\documentclass[sigconf, nonacm]{acmart}

\usepackage{hyperref}
\usepackage{url}
\usepackage{booktabs}
\usepackage{multirow,multicol}
\usepackage{graphicx}
\usepackage{xcolor}
\usepackage{float}
\usepackage{wrapfig}
\usepackage{soul}
\usepackage{tablefootnote}
\usepackage{pifont}
\usepackage{ulem}
\usepackage{tcolorbox}

\newcommand\vldbdoi{10.14778/3748191.3748222}
\newcommand\vldbpages{3655 - 3668}
\newcommand\vldbvolume{18}
\newcommand\vldbissue{10}
\newcommand\vldbyear{2025}
\newcommand\vldbauthors{\authors}
\newcommand\vldbtitle{\shorttitle} 
\newcommand\vldbavailabilityurl{URL_TO_YOUR_ARTIFACTS}
\newcommand\vldbpagestyle{empty} 

\newcommand{\cmark}{\ding{51}} % Checkmark symbol
\newcommand{\xmark}{\ding{55}} % Cross symbol

\begin{document}

\definecolor{mydarked}{rgb}{0.6, 0.1, 0.1} % Define a custom dark red color

% \definecolor{reviewer1}{rgb}{1.0, 0.0, 0.0} 

% \definecolor{reviewer2}{rgb}{0.0, 0.0, 1.0} 

% \definecolor{reviewer3}{rgb}{0.0, 0.7, 0.0}

% \definecolor{reviewer_shepherd}{rgb}{1.0, 0.3, 0.0}

\definecolor{reviewer1}{rgb}{0.0, 0.0, 0.0} 

\definecolor{reviewer2}{rgb}{0.0, 0.0, 0.0} 

\definecolor{reviewer3}{rgb}{0.0, 0.0, 0.0}

\definecolor{reviewer_shepherd}{rgb}{0.0, 0.0, 0.0}

\title{EVOSCHEMA: TOWARDS TEXT-TO-SQL ROBUSTNESS AGAINST SCHEMA EVOLUTION}

%%
%% The "author" command and its associated commands are used to define the authors and their affiliations.
\author{Tianshu Zhang}
\affiliation{%
  \institution{The Ohio State University}
  \streetaddress{P.O. Box 1212}
  \city{Columbus}
  \state{OH}
  \postcode{43017-6221}
}
\email{zhang.11535@osu.edu}

\author{Kun Qian}
\orcid{0000-0002-1825-0097}
\affiliation{%
  \institution{Adobe Inc.}
  \streetaddress{1 Th{\o}rv{\"a}ld Circle}
  \city{Seattle}
  \state{WA}
}
\email{kunq@adobe.com}

\author{Siddhartha Sahai}
\orcid{0000-0001-5109-3700}
\affiliation{%
\institution{Adobe Inc.}
  \streetaddress{1 Th{\o}rv{\"a}ld Circle}
  \city{Seattle}
  \state{WA}
}
\email{siddharthas@adobe.com}

\author{Yuan Tian}
\affiliation{%
  \institution{Purdue University}
  \city{West Lafayette}
  \state{IN}
  % \country{USA}
}
\email{tian211@purdue.edu}

\author{Shaddy Garg}
\affiliation{%
  \institution{Adobe Inc.}
  \city{Bangalore}
  \country{}
}
\email{shadgarg@adobe.com}

\author{Huan Sun}
\affiliation{%
  \institution{The Ohio State University}
  \city{Columbus}
  \country{OH}
}
\email{sun.397@osu.edu}

\author{Yunyao Li}
\affiliation{%
  \institution{Adobe Inc.}
  \city{San Jose}
  \country{CA}
}
\email{yunyaol@adobe.com}

\newcommand{\nop}[1]{}

\begin{abstract}

Neural text-to-SQL models, which translate natural language questions (NLQs) into SQL queries given a database schema, have achieved remarkable performance. However, database schemas frequently evolve to meet new requirements. Such schema evolution often leads to performance degradation for models trained on static schemas. Existing work either mainly focuses on simply paraphrasing some syntactic or semantic mappings among NLQ, DB and SQL, or lacks a comprehensive and controllable way to investigate the model robustness issue under the schema evolution, which is insufficient when facing the increasingly complex and rich database schema changes in reality, especially in the LLM era.

\nop{In this work, we approach this critical problem by introducing a novel framework, \texttt{EvoSchema}, to systematically simulate diverse schema changes that occur in real-world scenarios. \textsf{EvoSchema} builds on our newly defined schema evolution taxonomy, which encompasses a comprehensive set of ten perturbation types, covering both column-level and table-level modifications. We utilize this framework to build an evaluation benchmark to assess the models' robustness against different schema evolution types. Meanwhile, we propose a new training paradigm, which augments existing training data with diverse schema designs and \nop{uses this framework to augment the existing training data with different schema evolution types under the same questions, and enhances the model robustness by}forces the model to distinguish the schema difference for the same questions to avoid learning spurious patterns. Our experiments demonstrate that the existing models are more easily affected by table-level perturbations than column-level perturbations. In addition, the models trained under our paradigm exhibit significantly improved robustness, achieving up to 33 points improvement on the evaluation benchmark compared to models trained on unperturbed data. This work represents a significant step towards building more resilient text-to-SQL systems capable of handling the dynamic nature of database schemas.\footnote{Our code and data will be publicly available.}}

To address the challenges posed by schema evolution, we present \texttt{EvoSchema}, a comprehensive benchmark designed to assess and enhance the robustness of text-to-SQL systems under real-world schema changes. \texttt{EvoSchema} introduces a novel schema evolution taxonomy, encompassing ten perturbation types across column-level and table-level modifications, systematically simulating the dynamic nature of database schemas. Through \texttt{EvoSchema}, we conduct an in-depth evaluation spanning different open-source and closed-source LLMs, revealing that table-level perturbations have a significantly greater impact on model performance compared to column-level changes. Furthermore, \texttt{EvoSchema} inspires the development of more resilient text-to-SQL systems, in terms of both model training and database design. The models trained on \texttt{EvoSchema}'s diverse
schema designs can force the model to distinguish the schema difference for the
same questions to avoid learning spurious patterns, which demonstrate remarkable robustness compared to those trained on unperturbed data on average. This benchmark offers valuable insights into model behavior and a path forward for designing systems capable of thriving in dynamic, real-world environments.

% The abstract paragraph should be indented 1/2~inch (3~picas) on both left and
% right-hand margins. Use 10~point type, with a vertical spacing of 11~points.
% The word \textsc{Abstract} must be centered, in small caps, and in point size 12. Two
% line spaces precede the abstract. The abstract must be limited to one
% paragraph.
\end{abstract}

% \begin{abstract}
% Praesent imperdiet, lacus nec varius placerat, est ex eleifend justo, a vulputate leo massa consectetur nunc. Donec posuere in mi ut tempus. Pellentesque sem odio, faucibus non mi in, laoreet maximus arcu. In hac habitasse platea dictumst. Nunc euismod neque eu urna accumsan, vitae vehicula metus tincidunt. Maecenas congue tortor nec varius pellentesque. Pellentesque bibendum libero ac dignissim euismod. Aliquam justo ante, pretium vel mollis sed, consectetur accumsan nibh. Nulla sit amet sollicitudin est. Etiam ullamcorper diam a sapien lacinia faucibus.
% \end{abstract}

\maketitle

\pagestyle{\vldbpagestyle}
\begingroup\small\noindent\raggedright\textbf{PVLDB Reference Format:}\\
\vldbauthors. \vldbtitle. PVLDB, \vldbvolume(\vldbissue): \vldbpages, \vldbyear.\\
\href{https://doi.org/\vldbdoi}{doi:\vldbdoi}
\endgroup
\begingroup
\renewcommand\thefootnote{}\footnote{\noindent
This work is licensed under the Creative Commons BY-NC-ND 4.0 International License. Visit \url{https://creativecommons.org/licenses/by-nc-nd/4.0/} to view a copy of this license. For any use beyond those covered by this license, obtain permission by emailing \href{mailto:info@vldb.org}{info@vldb.org}. Copyright is held by the owner/author(s). Publication rights licensed to the VLDB Endowment. \\
\raggedright Proceedings of the VLDB Endowment, Vol. \vldbvolume, No. \vldbissue\ %
ISSN 2150-8097. \\
\href{https://doi.org/\vldbdoi}{doi:\vldbdoi} \\
}\addtocounter{footnote}{-1}\endgroup

\ifdefempty{\vldbavailabilityurl}{}{
\vspace{.3cm}
\begingroup\small\noindent\raggedright\textbf{PVLDB Artifact Availability:}\\
The source code, data, and/or other artifacts have been made available at \url{https://github.com/zhangtianshu/EvoSchema}.
\endgroup
}

\section{introduction}

Text-to-SQL parsing aims to translate natural language questions (NLQs) into SQL queries given a database schema, enabling the development of natural language interfaces that allow users to query data and invoke services without requiring programming skills \citep{wang-etal-2020-rat, zhang2024benchmarkingtexttosqlcapabilitylarge, yu-etal-2018-spider, zhang2023federatedlearningsemanticparsing, bird, tai-etal-2023-exploring}. Existing neural text-to-SQL models have achieved remarkable performance on existing benchmarks \citep{bird, yu-etal-2018-spider}, which play an important role in empowering different platforms such as business and marketing platforms \citep{song2024financial, zhang2024finsql} and being integrated into virtual assistants to enable real-time data query and analysis \citep{deksne2022virtual}.

\begin{figure*}[t]
\begin{center}
%\framebox[4.0in]{$\;$}
% \fbox{\rule[-.5cm]{0cm}{4cm} \rule[-.5cm]{4cm}{0cm}}
\includegraphics[width=0.85\linewidth]{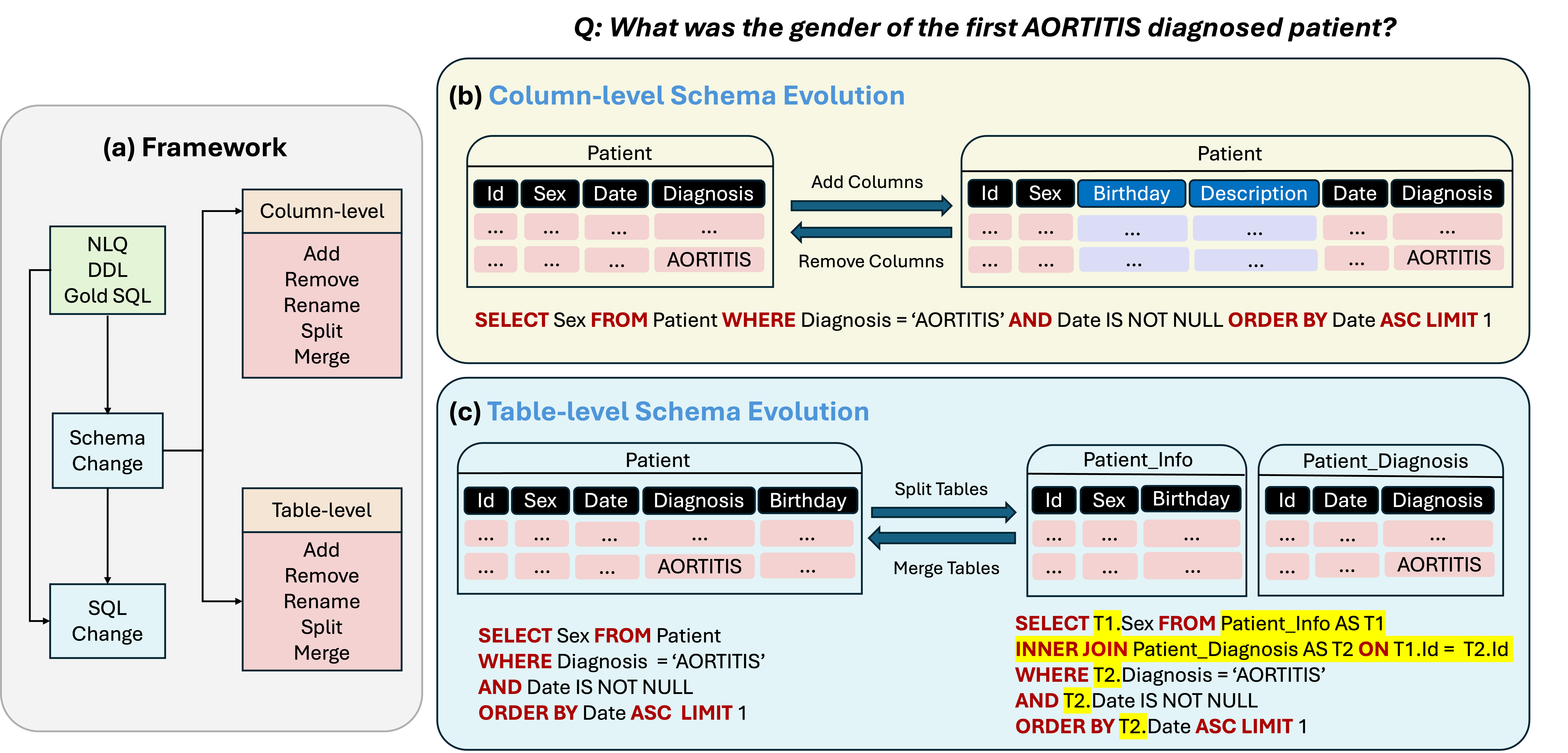}
\end{center}
\caption{The left (a) is the overview of the framework to collect \texttt{EvoSchema} dataset. The top right (b) is a column-level schema evolution example; the bottom right (c) is a table-level schema evolution example.\nop{\hs{might index the subfigures as (a/b/c) so in intro we could refer to them.}}}
\label{fig:overview}
\end{figure*}

However, database schemas are not static; they frequently evolve to accommodate new use cases and improve efficiency \citep{hillenbrand2021managing, cleve2015understanding}. For instance, depending on the scenario, a large patient table might be merged from or split into two tables: a patient information table and a patient diagnosis table (Figure \ref{fig:overview}-c), to reduce redundancy, enhance data integrity, and optimize performance \citep{Database_normalization_design_pattern}. Such schema evolution occurs frequently, which often leads to distribution shifts \citep{distributionalshiftsml, koh2021wilds} such as nomenclature shifts, data granularity shifts, table and column relation shifts and schema complexity shifts. These distribution shifts can cause significant performance degradation when the model trained on old database schema is adapting to new schema designs.

This challenge highlights a critical issue in model robustness: how well can a text-to-SQL model adapt to changes in the database schema? Recent studies introduce evaluation benchmarks designed to expose robustness issues by perturbing NLQs, databases or SQL queries \citep{chang2023drspider, Deng_2021, pi-etal-2022-towards, mt-teql}. However, these studies have at least one of the following limitations: 1) mainly focus on the syntactic paraphrasing or simple semantic mappings among NLQ, DB and SQL \citep{chang2023drspider, Deng_2021}; (2) lack a taxonomy of comprehensive schema evolution types \citep{pi-etal-2022-towards}; (3) only focus on schema evolution that does not lead to SQL changes \citep{mt-teql}.
\nop{tend to focus on syntactic paraphrasing or simple semantic mappings, such as different representations of numbers or name abbreviations across NLQ, DB, and SQL. While some work analyzes schema changes, they mainly focus on irrelevant column modifications that do not affect SQL \citep{mt-teql} or with limited perturbation types \citep{pi-etal-2022-towards}.}These efforts are insufficient in the face of increasingly complex and rich database schema changes found in reality.\nop{Moreover, the advent of LLMs has mitigated many linguistic challenges, further emphasizing the need for robust adaptation to structural changes in database schemas.  (TODO: add robust training \& add citations for several other perturbations)}
Meanwhile, while it is natural to consider collecting new data after schema evolution for retraining a model, repeating the entire model training life cycle frequently can be costly in terms of both time and resources.

\nop{\yuancomment{I think this paragraph is a bit detailed (like the sentences in related work). I feel we can summarize existing works and focus on discussing why they are insufficient---They can't answer the following 2 research questions. Then to answer the questions, we build a new, more comprehensive framework. This framework's key feature is to simulate real changes by perturbations on BIRD \citep{bird}. Here we should somehow discuss why perturbation works. For example, it not only augments the dataset with more types of schemas but also allows models to learn the structure difference within a schema. These types of data are not enough in existing datasets (novelty). In fact, we find perturbations help a lot (contribution).}}

\begin{figure*}[ht]
\begin{center}
%\framebox[4.0in]{$\;$}
% \fbox{\rule[-.5cm]{0cm}{4cm} \rule[-.5cm]{4cm}{0cm}}
\includegraphics[width=0.85\linewidth]{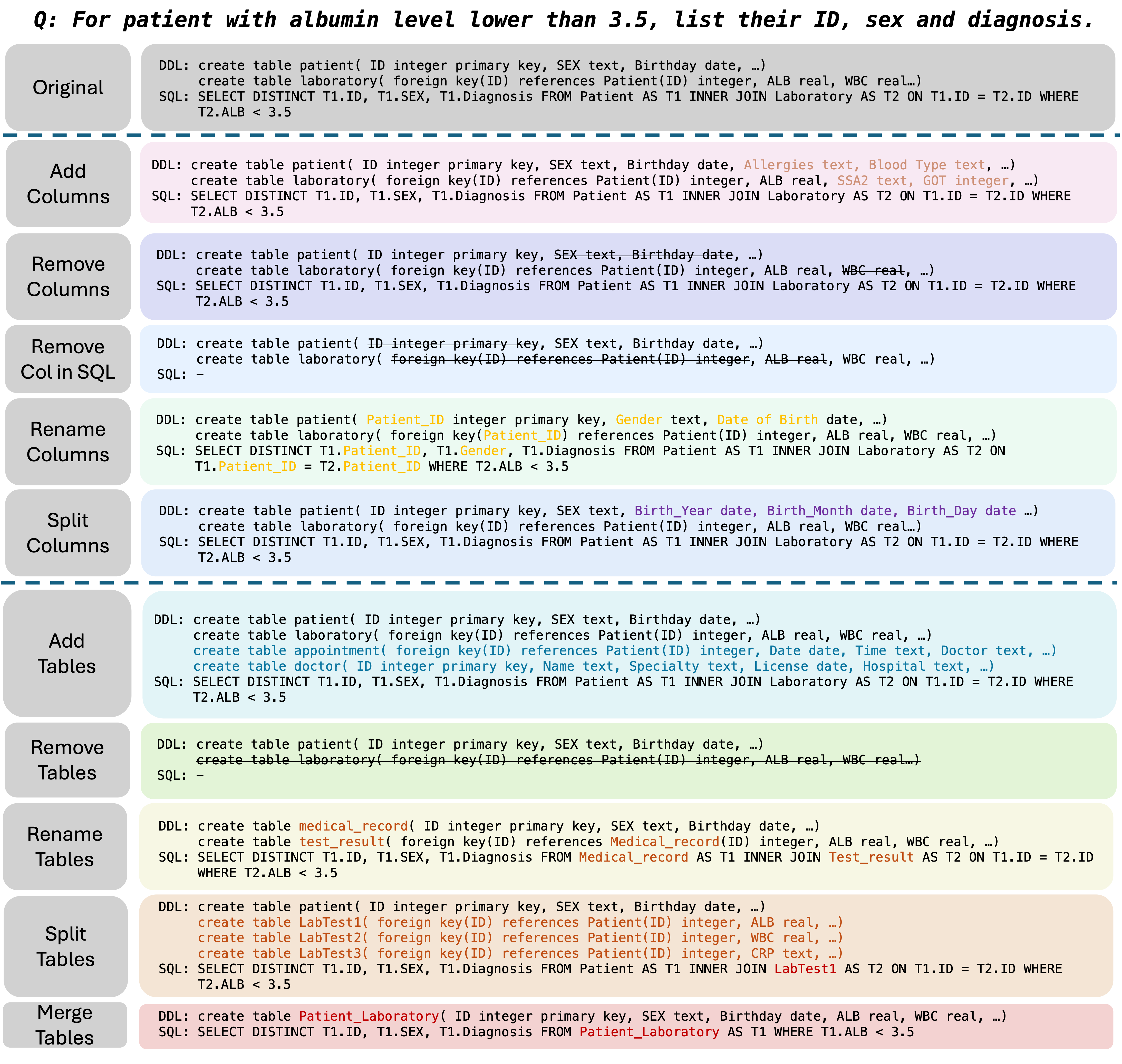}
\end{center}
\caption{An overview of different perturbation types of \texttt{EvoSchema}. The top is an unperturbed example in BIRD \citep{bird}; the middle is the column-level perturbation; the bottom is the table-level perturbation. ``Remove Col in SQL": remove columns that appear in gold SQL; ``Remove Tables": the relevant tables appear in gold SQL are removed. Thus there is no gold SQL for these two cases. Note we don't illustrate ``Merge Columns'' in the figure as this example is not suitable for applying merging column changes.}
\label{fig:appendix_examples}
\end{figure*}

Under this background, we seek to answer the following two questions: (1) How sensitive are existing text-to-SQL models to various types of database schema changes? (2) How can we train a more robust text-to-SQL model that not only performs well on existing database schemas but also adapts effectively to schema changes? Towards this end, we introduce \texttt{EvoSchema}, a new dataset that covers a wide range of realistic schema design changes by perturbations on BIRD \citep{bird}. As illustrated in Figure \ref{fig:overview} and Figure \ref{fig:appendix_examples}, \texttt{EvoSchema} builds upon our newly defined taxonomy, which encompasses a total of ten types of perturbations over schema, covering both column-level and table-level changes. Column-level perturbations include adding, removing, renaming, splitting and merging columns, while table-level perturbations involve adding, removing, renaming, splitting, and merging tables. We keep the NLQs fixed and examine the robustness of a model under different granularities of schema evolution, and show that existing models are more easily affected by table-level perturbations than column-level perturbations.

\nop{which not only builds the foundation to evaluate the robustness against different granularities of schema evolution, but also provides insights to improve models' ability by forcing models to distinguish the structure difference within the schema so as to avoid learning the spurious patterns.

\nop{\yuancomment{There is a logic gap in why we want to establish a general principle.}
To establish a general principle for building more robust text-to-SQL parsing systems, we propose a schema evolution synthesis framework, \texttt{EvoSchema}. This framework simulates a wide range of realistic schema design changes, generating data that serves both for robustness evaluation and for training models to better withstand schema evolution. 
\yuancomment{We can refer to Figure 1}}

As illustrated in Figure \ref{fig:overview}, \texttt{EvoSchema} framework builds upon our newly defined taxonomy, which encompasses a total of ten types of perturbations over schema, covering both column-level and table-level changes. Column-level perturbations include adding, removing, renaming, splitting and merging columns, while table-level perturbations involve adding, removing, renaming, splitting, and merging tables. We keep the NLQs fixed and examine the robustness of a model under different schema evolutions, and show that existing models are more easily affected by table-level perturbations than column-level perturbations\nop{\hs{only by table-level perturbations? not so much affected by column-level perturbations?}}.}

Moreover, the training set in \texttt{EvoSchema} can be used to enhance models' robustness. The models can be trained with the same questions but coupled with different schema designs to generate the corresponding SQL queries. This training procedure forces the model to distinguish the schema difference which can help models gain a stronger ability to recognize the correct table and column relation and map them to the questions. Our experimental results demonstrate that the perturbation training data in \texttt{EvoSchema} can help train better text-to-SQL models, which are more robust to different schema evolution types on average, especially on table-level perturbations\nop{\hs{why specifically emphasize on table-level perturbations (both here and earlier)?}}.
\nop{\yuancomment{I think we can highlight that the perturbation data generated by this framework can help improve training better text-to-SQL models}
By synthesizing training data through our framework, we train text-to-SQL models and evaluate their performance across different schema perturbation types, demonstrating significant improvements in handling table-level perturbations compared to models trained without such perturbations.}

In summary, our main contributions are as follows:
\begin{itemize}
  \item We formulate a critical schema evolution adaptive text-to-SQL problem and present a new dataset, \texttt{EvoSchema} to study this problem. We introduce a comprehensive taxonomy of the schema evolution types and build the datasets based on the taxonomy to get realistic schema designs by column-level and table-level perturbations on BIRD.
  \item We conduct thorough and comprehensive assessment of model robustness against various schema perturbations spanning different open-source and closed-source LLMs on our evaluation benchmark, and find that table-level perturbations have a significantly greater impact on model performance compared to column-level changes. Besides, we introduce two evaluation metrics: Table Match F1 and Column Match F1, to rigorously evaluate the performance of text-to-SQL models under schema evolution scenarios and provide fine-grained insights into model robustness.
  \item Our constructed training set inspires a new training paradigm: augmenting the existing training data with different schema designs, which not only increase the data diversity, but also force the model to distinguish the schema difference during training. Our approach yields better text-to-SQL models that achieve up to 33 points gain on different types of schema perturbation evaluation data, compared to models trained on unperturbed, original training data.
\nop{\yuancomment{what metric and benchmark.}}
\end{itemize}

\section[Related Work]{Related Work}
\label{related_work}
\noindent \textbf{Robustness in Text-to-SQL:}
Existing research on text-to-SQL robustness is mainly
two-fold: robustness evaluation and robustness training. Recent studies introduce evaluation benchmarks designed to expose robustness issues by perturbing NLQs, databases or SQL queries. However, these studies tend to focus on syntactic paraphrasing or simple semantic mappings, such as
different representations of numbers or name abbreviations across NLQ, DB, and SQL \citep{chang2023drspider, Deng_2021}. While
some work analyzes schema changes, they mainly focus on irrelevant column modifications that do
not affect SQL \citep{mt-teql} or with limited perturbation types \citep{pi-etal-2022-towards}. These efforts are insufficient in the face of increasingly complex and rich database schemas found in modern datasets. \textcolor{reviewer1}{Though FootballDB \citep{footballdb} tackles a similar schema design problem for better SQL written, they focus on reducing multiple foreign key mappings among tables and reducing the JOIN paths in the SQL. Different from theirs, we tackle the schema evolution problem, which is not only for the schema design on the existing data, but also needs to consider how new data and information will change the schema design. Besides, we approach it through a different angle, where our scheme design contains 10 column-level and table-level changes. And our provided schema evolution framework allows us to try different schema design on multiple databases to get more generalizable findings, while FootballDB \citep{footballdb} can only support the exploration on a single database.} Moreover, the advent of LLMs has mitigated many linguistic challenges, further emphasizing the need for robust adaptation to structural changes in database
schemas. For robust training, existing methods employ strategies like decomposing tasks so that models generate each sub-clause individually before merging them \citep{gao-etal-2022-towards-generalizable}, or using execution-guided decoding to eliminate incorrect sub-clauses \citep{wang2018robusttexttosqlgenerationexecutionguided}. While these approaches focus on enhancing various aspects of text-to-SQL robustness, our work specifically addresses the challenge of schema evolution.

\nop{\noindent \textbf{LLMs for Text-to-SQL:}
Most recently, the LLM-based approaches
for text-to-SQL are mainly two-fold: in-context learning \citep{zhang2023actsql, tai-etal-2023-exploring, gao2023text, li2024llm, li2024dawn} and finetuning \citep{li2024codes, zhuang2024structlm, li2024llm, li2024dawn}. The former prompts proprietary LLMs such as GPT series \footnote{https://platform.openai.com/docs/models} and Claude \footnote{https://www.anthropic.com/news/claude-3-family} for SQL generation, while the latter directly finetunes open-source LLMs on text-to-SQL datasets. These models are designed for question understanding, schema comprehension and SQL generation, which have achieved remarkable performance on the existing open benchmarks \citep{bird, yu-etal-2018-spider}. In our work, we leverage both the proprietary LLMs such as GPT-3.5 and GPT-4 \citep{openai2024gpt4technicalreport} and open-source LLMs such as Code Llama \citep{rozière2024codellamaopenfoundation}, Mistral \citep{jiang2023mistral7b}, Llama 3 \citep{dubey2024llama3herdmodels} and SQLCoder \footref{myfootnote} to explore the models' robustness against schema evolution.}

\noindent \textbf{LLMs for Text-to-SQL:} 
Most recently, the LLM-based approaches
for text-to-SQL are mainly two-fold: in-context learning \citep{zhang2023actsql, tai-etal-2023-exploring, gao2024text, li2024llm, li2024dawn} and finetuning \citep{li2024codes, zhuang2024structlm, li2024llm, li2024dawn}. The former prompts proprietary LLMs such as GPT series \footnote{https://platform.openai.com/docs/models} and Claude \footnote{https://www.anthropic.com/news/claude-3-family} for SQL generation without additional model training, while the latter involves adapting open-source LLMs to text-to-SQL datasets, tailoring these models directly to the task through supervised learning. These models are designed for question understanding, schema comprehension and SQL generation, which have achieved remarkable performance on the existing open benchmarks \citep{bird, yu-etal-2018-spider}. \nop{In our work, we leverage both the proprietary LLMs such as GPT-3.5 and GPT-4 \citep{openai2024gpt4technicalreport} and the cutting-edge open-source LLMs such as Code Llama \citep{rozière2024codellamaopenfoundation}, Mistral \citep{jiang2023mistral7b}, Llama 3 \citep{dubey2024llama3herdmodels} and SQLCoder \footref{myfootnote} to explore the models' robustness against schema evolution.} \textcolor{reviewer_shepherd}{\citet{liu2024survey} provides a comprehensive review of the NL2SQL lifecycle, covering models, benchmarks, data synthesis, evaluation, and error analysis. While it identifies schema variation as a challenge, it does not explore it in depth. Our work focuses specifically on schema evolution robustness by evaluating recent and powerful LLMs (e.g., Code Llama, Mistral, SQLCoder, LLaMA 3, GPT-3.5, GPT-4) without preprocessing or postprocessing. We introduce \texttt{EvoSchema}, a benchmark with controlled schema perturbations that guides both evaluation and structured training data synthesis. In addition to standard execution accuracy and human evaluation, we propose two fine-grained metrics: Table Match F1 and Column Match F1 that directly reflect our table-level and column-level perturbation taxonomy. \citet{li2024dawn} evaluates LLMs on unperturbed Spider and BIRD datasets and also experiments on natural language variation but keep schema and SQL fixed; in contrast, our work systematically varies the schema while keeping the natural language fixed.\nop{\citet{liu2024survey} systematically reviews the entire NL2SQL lifecycle across four dimensions: NL2SQL models, benchmark, evaluation and error analysis. While schema generalization is identified as a challenge, it is not the central focus of their empirical discussion. Our work complements this by providing a systematic and empirical framework—including a taxonomy of schema perturbations and structured training data synthesis—to specifically study this issue. \citet{li2024dawn}  benchmark both proprietary and open-source LLMs on unperturbed datasets: Spider and BIRD. Their query variance testing investigates model robustness to paraphrased natural language queries, holding the schema fixed. Our work is orthogonal: we fix the natural language input and systematically vary the database schema, enabling a different and complementary evaluation of model generalization.}}

\nop{Instruction tuning in large language models (LLMs) using \textless instruction, input, output\textgreater pairs is a crucial technique for enhancing the capabilities and controllability of these models. By providing targeted instructions, models can be guided to produce outputs that align with specific response characteristics or domain knowledge, facilitating rapid adaptation to particular domains without the need for extensive retraining or architectural modifications. To address diverse needs, various instruction tuning datasets have been developed to refine LLMs' behavior. While open-source models like CodeLlama, Mistral, and LLaMA 3 have shown strong performance in code generation tasks, they are not tailored to the specific demands of text-to-SQL scenarios, particularly those involving schema evolution. Given the unique challenges of schema evolution in text-to-SQL tasks, we perform instruction tuning on these base models using our dataset. This approach allows us to customize the models to better handle schema changes, ensuring they are equipped to tackle the complexities of our specific problem domain.}

% LLMs using \textless instruction, input, output\textgreater pairs in a supervised fashion is a crucial technique to enhance the
% capabilities and controllability of LLMs. The instructions serve to constrain the model’s outputs to align with the desired response characteristics or domain knowledge and can help LLMs
% rapidly adapt to a specific domain without extensive retraining or architecture designs. Therefore, different instruction tuning datasets have been proposed to guide LLMs’ behaviors. There are some well-known open-source base models such as codellama, mistral, and llama3 that are good at code generation tasks. However, they are not customized to our text-to-SQL cases. In order to make the model better fit our data format, we do instruction tuning on these base models using our dataset to explore our schema evolution problem.

% SQLCoder XX specialist open-source models.
\section{EvoSchema Dataset}

\subsection{Background}

In the dynamic landscape of databases, schemas frequently evolve to meet new demands, introducing significant challenges for text-to-SQL models \citep{Relational_Database_Schema_Evolution:_An_Industrial_Case_Study, cleve2015understanding}. These schema changes can vary widely, from minor modifications to complete restructuring, and can significantly impact the performance of models trained on static schemas. \nop{To address this challenge, our study focuses on developing a framework that comprehensively covers potential schema evolution types, thereby fostering both robustness evaluation and inspiring robust model training for text-to-SQL systems.}In realistic scenarios, a database can often contain a large number of tables, and only several related tables are responsible for a natural language question (NLQ). In our experiment, we represent the relevant database schema using Data Definition Language (DDL) \footnote{DDL defines the structure and properties of a database, providing
detailed information necessary for database creation, including column types and
primary/foreign keys.} and combine it with the NLQ as input. This input is then used to prompt the model to generate the corresponding SQL query.

\subsection{Rationale for Schema Evolution Types} 
When a database schema evolves, it can induce distribution shifts in the data that may impact model performance. We categorize potential distribution shifts into four types: nomenclature shifts, data granularity shifts, table and column relation shifts, and schema complexity shifts. (1) Nomenclature shifts occur when tables and columns are renamed, which may alter the convention of the established terminology within the schema. For example, tables originally named ``Products", ``Customers", and ``Orders" might be renamed to ``Items", ``Clients", and ``Purchases", respectively. Such changes often reflect updates in business terminology or compliance with new standards. A desired model should handle those nomenclature shifts to adapt to the new terminology. (2) Data granularity shifts arise from adding or removing columns or tables, which changes the level of detailedness captured in the database. For instance, an ``Employee" table with a single ``ContactNumber" field might involve another two separate ``WorkContact" and ``PersonalContact" fields later. This increases the data granularity to meet new requirements, necessitating models to adapt to more complex and detailed semantics. (3) Table and column relation shifts and schema complexity shifts mainly result from restructuring tables through splitting or merging. This process can highly affect how each table is related to other tables by which column. Both the primary keys and foreign keys may change along with the table restructure. Besides, the schema complexity may change when multiple tables merge from or split into one table. A desired model is expected to be robust to such changes. By categorizing the distribution shifts caused by schema evolution, we can more effectively understand and evaluate a model's capacity to adapt to changes in the underlying database schema.

\nop{By understanding and categorizing these types of distributional shifts caused by potential schema evolution, we can better assess a model's ability to handle distributional shifts resulting from changes in the underlying data schema.}

\subsection{Schema Evolution Synthesis Framework}

\begin{figure*}[t]
\begin{center}
%\framebox[4.0in]{$\;$}
% \fbox{\rule[-.5cm]{0cm}{4cm} \rule[-.5cm]{4cm}{0cm}}
\includegraphics[width=0.7\linewidth]{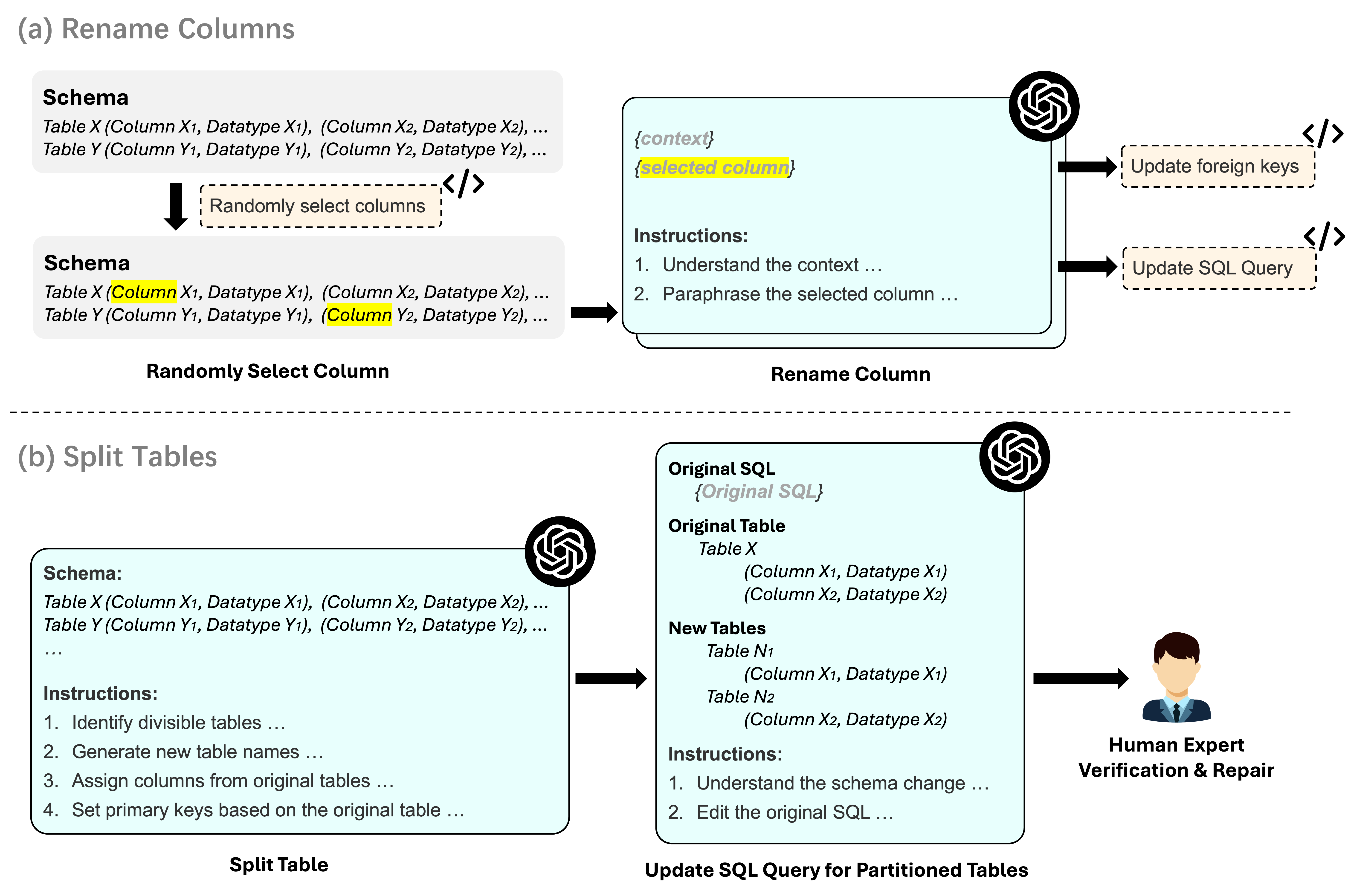}
\end{center}
\caption{This figure shows two examples of our data collection procedure of \texttt{EvoSchema}. The top (a) is a ``rename columns" data collection procedure; the bottom (b) is a ``split tables" data collection procedure. The blue box indicates prompting GPT models for the generation. ``</>" means programmatically processing the data.}
\label{fig:data_collection}
\end{figure*}

Our study aims to cover comprehensive potential schema evolution types, which can foster the robustness evaluation of the existing text-to-SQL models and inspire robust model training. We synthesize all the schema evolution types through hybrid strategies, which will leverage both the heuristic rules to guarantee the data quality and LLMs to ensure diversity.

\noindent \textbf{Broad Coverage of Different Schema Evolution Types:} We aim to encapsulate a broad range of schema evolution types, recognizing their prevalence and impact in real-world scenarios. Specifically, our schema evolution taxonomy includes both column-level and table-level perturbations, which are categorized into ten distinct types. Column-level perturbations comprise five types: adding, removing, renaming, splitting and merging columns, where modifications are restricted to the columns within existing tables.\nop{For example, a column-level scenario might involve adding a new ``discount" column to a sales table, which could affect how the SQL queries calculate total sales.} Table-level perturbations encompass five types: adding, removing, renaming, splitting, and merging tables. \nop{which necessitate managing relationships among different tables and can substantially alter data distributions.For example, a table-level scenario might involve merging two tables, such as combining a ``customer" table with an ``order details" table into a single ``customer\_orders" table, to streamline database structure and queries.} These perturbations occur frequently in practice, underscoring the need for text-to-SQL models that can robustly handle such changes.

% We aim to cover diverse and comprehensive possible schema design changes. Our schema evolution types cover both column-level and table-level perturbations, which are eight types in total. Column-level perturbations include three types: adding, removing and renaming columns, which only revise the columns of the existing tables, while table-level perturbations include five types: adding, removing, renaming, splitting and merging tables, which need to handle the relation among different tables and will largely cause the data distribution change as well. These schema evolutions are highly possible in the real world and occur very frequently, which signifies the importance of a robust text-to-SQL model to work well on these changes.

% (can add a table to show the data statistics, can put it into the experiment section)

\noindent \textbf{Hybrid Data Synthesis Strategies:}
% To ensure both the diversity and quality of the synthesized schema perturbations, we combine heuristics rules and LLMs to synthesize different perturbation types. For all the given seed \textless NLQ, relevant schema, SQL\textgreater  triple, we keep the NLQ fixed for all the perturbation types and only revise the relevant schema and make sure the SQL changes accordingly along with the database schema change if needed. For adding columns, we prompt GPT-3.5 to synthesize the columns that are suitable in the current context for each table in the relevant schema and randomly insert them into the table. For removing columns, we randomly remove columns in the given table schema but not appear in the gold SQL. For both adding columns and removing columns, the NLQ and the gold SQL are fixed. For renaming columns, we randomly select columns and use GPT-3.5 to synthesize the similar name and replace them. 
To ensure both diversity and quality in the generation of schema perturbations, we employ a combination of heuristics and GPT models to synthesize various perturbation types. For each given seed instance in BIRD \citep{bird}, consisting of a \textit{\textless NLQ, relevant schema, SQL\textgreater} triple, we maintain the natural language question (NLQ) fixed across all perturbation types, while only modifying the relevant schema. The corresponding SQL query is adjusted as necessary to remain consistent with the changes in the database schema.

\subsection{Seed Dataset Selection} 
\label{seed_data_select}
For building \texttt{Evoschema} benchmark, we utilize the BIRD \citep{bird} dataset as the seed data, which is specifically designed for the text-to-SQL task. Compared to Spider \citep{yu-etal-2018-spider}, which is commonly used to study text-to-SQL robustness, BIRD features more intricate, realistic, and extensive databases, as well as more complex SQL queries that include keywords often missing in Spider. BIRD consists of NLQs, corresponding database schemas, and gold SQL queries and encompasses a wide range of real-world database scenarios, which provides a robust foundation for evaluating the performance of models in translating NLQs into SQLs.

\textcolor{reviewer1}{Schema Perturbations: To evaluate the robustness of the text-to-SQL models, \texttt{EvoSchema} not only includes the BIRD dataset in their original form but also augmented it with various column-level and table-level schema perturbations. We ensure that the NLQs remain fixed, while the schema and SQL queries are adjusted as necessary to reflect the changes introduced by our perturbations. We follow the standard train/dev split provided with BIRD, and apply all the perturbations on both training data and evaluation data. The data statistics of \texttt{EvoSchema} are in Table \ref{tab:data_stat_plus_manipulated_stat} and the examples of different perturbation types are in Figure \ref{fig:appendix_examples}.}

\subsection{Data Generation}

We design a framework to simulate different types of schema perturbations in a configurable way. For adding or renaming columns, both the modified column size and the column position in the tables are set randomly, and we set the original column size in the table as the maximum number of columns to be changed. For removing columns, we can randomly remove important or unimportant columns from the existing relevant tables. The important columns are the columns that appear in the gold SQL, which will inevitably affect the prediction. For adding, removing, or renaming tables, we randomly add, remove or rename one or multiple tables.

% When the schema evolves, the model will experience the distribution shift of the data. We define the potential schema evolution types that have different affects on the distribution shifts in terms of nomenclature shifts, data granularity shifts, table and column relation shifts and schema complexity shifts.
% Renaming tables and columns can lead to nomenclature shifts which is how the tradition of defining the terminology changes. For example, 
% Before: Consistent names like Products, Customers, Orders. After: Names change to Items, Clients, Purchases.  Nomenclature shifts to reflect new business terminology or standards. In such cases, we evaluate whether the model can handle the linguistic mappings well. Adding or removing columns can lead to data granularity shifts. Before: A single Address field in Customers. After: Separate BillingAddress and ShippingAddress fields added. Data granularity in the database increases or decreases to meet new requirements. Restructuring the tables such as splitting or merging tables can lead to table and
% column relation shifts and schema complexity shifts
% Before: Highly normalized schema with multiple related tables. After: Denormalized schema combining tables for performance. Adjusting schema complexity to balance between speed and storage efficiency.

\noindent\textbf{Schema Change:} To ensure the diversity and reasonability of the synthesized schema, we leverage the capabilities of GPT-3.5 and GPT-4 to synthesize realistic and contextually appropriate columns or tables, which help effectively produce high-quality synthetic data that meets our requirements. For adding or renaming columns and tables, we input the existing relevant tables to GPT-3.5, and let the model generate the potential tables or columns that fit the context. For splitting tables or merging tables, since they are more complex than other perturbations, we use GPT-4 to choose the tables that can be split or merged and then use the modified tables to replace the original ones. For adding or renaming columns and tables, we apply heuristics to filter out the repeated ones in the synthesized tables or columns. Besides, to ensure the correct relationship among different tables after modifying the schema, we apply heuristics to ensure all the foreign keys change along with their referenced table names and column names. When removing columns or tables, any foreign keys in other tables that reference the removed columns or tables will be removed as well.

\noindent\textbf{SQL Change:} To ensure the consistency of the \textit{\textless NLQ, relevant schema, SQL\textgreater}, after we change the relevant table schema,
we revise the gold SQL accordingly. Since the NLQs are the same for adding or removing columns and tables, and the schema evolution here doesn't affect answering the questions, we keep the gold SQL unchanged for these perturbation types. For renaming columns or tables, we revise gold SQL if they appear in the gold SQL. For table splitting or merging, due to the complexity and variation in the required SQL changes, we use GPT-4 to revise the gold SQL\nop{\hs{potential question: how good is the quality? do you use the revised SQL to evaluate models or just for training?}}. This revision is based on the mappings from the original to the new tables and columns, as well as the necessary adjustments to the JOIN paths. We manually check the edited gold SQL for the evaluation benchmark to make sure they are correct.

\nop{By employing these strategies, \texttt{EvoSchema} offers a comprehensive and diverse set of schema evolution scenarios that mirror the complexities encountered in real-world database management. By integrating heuristics with LLM-generated perturbations followed by a human verification process, we maintain both of the diversity and quality, ensuring that the synthesized data is both realistic and challenging. }\nop{Our framework is configurable to generate different perturbation types, which enables us to systematically evaluate the impact of schema evolution on text-to-SQL model performance in a structured and controlled manner. Besides, \nop{\hs{I think you might start a new paragraph to elaborate on this sentence?}}this approach also lays the groundwork for improving the robustness of text-to-SQL models.}

\begin{table*}[ht]
\caption{Statistics of \texttt{EvoSchema} compared with existing benchmarks. ``Tab'': tables; ``DB'': database; ``Col'': columns; ``PK'': primary keys; ``FK'': foreign keys.}

\small
\resizebox{1.0\linewidth}{!}{
\centering
\begin{tabular}{@{}l|cccccccccc@{}}
\toprule
 \multirow{2}{*}{Perturbation Data} & \multirow{2}{*}{Column-level} & \multirow{2}{*}{Table-level} & Schema Evolution & \multirow{2}{*}{Multiple DB} & \multirow{2}{*}{Seed Data} & \multicolumn{5}{c}{Features of Seed Data (Average)} \\
 & & & Affects SQL & & & Tab/DB & Col/DB & Col/Tab & PK/DB & FK/DB \\
\midrule
\textcolor{reviewer1}{FootballDB \citep{footballdb}} & \textcolor{reviewer1}{-} & \textcolor{reviewer1}{reduce PK/FK references, reduce JOIN paths} & \textcolor{reviewer1}{\cmark} & \textcolor{reviewer1}{\xmark} & \textcolor{reviewer1}{FIFA World Cup \citep{fifa}} & \textcolor{reviewer1}{15} & \textcolor{reviewer1}{107} & \textcolor{reviewer1}{7.1} & \textcolor{reviewer1}{-} & \textcolor{reviewer1}{16}
\\

Dr.Spider \citep{chang2023drspider} & Rename & \xmark & \cmark & \cmark & Spider \citep{chang2023drspider} & 5.1 & 22.1 & 5.4 & 3.7 & 3.2 \\
ADVETA \citep{pi-etal-2022-towards} & Add; Rename & \xmark & \cmark & \cmark & Spider \citep{chang2023drspider} & 5.1 & 22.1 & 5.4 & 3.7 & 3.2\\ 
MT-TEQL \citep{mt-teql} & Add; Remove; Shuffle; Rename & Split; Merge; Shuffle & \xmark & \cmark & Spider \citep{chang2023drspider} & 5.1 & 22.1 & 5.4 & 3.7 & 3.2\\
EvoSchema (Ours) & Add; Remove; Rename; Split; Merge & Split; Merge; Rename; Add; Remove & \cmark & \cmark & BIRD \citep{bird} & 7.3 & 72.5 & 10.6 & 6.5 & 9.3  \\ 

\bottomrule
\end{tabular}
}
\label{tab:comparison_with_existing_benchmarks}
\end{table*}

\subsection{Data Collection of Each Perturbation Type}
\label{data_collection_for_each_type}
We first define heuristics for different perturbation types, then combine both GPT models' generation ability and programming to collect the data. Finally, we incorporate a human verification stage to control the data quality. Here are some general heuristics we should consider to maintain consistency and avoid conflicts when manipulating data: 1) Preserve Meaning: For renaming, the new column or table name should reflect the same meaning as the original name to avoid semantic confusion. 2) Avoid Conflicts: Ensure that the new column or table name does not conflict with existing column or table names within the same or other tables in the database. 3) Update References: Update all references to the new column or tables in foreign keys in other tables. 4) Revise SQL: Update all SQL queries referencing the new columns or tables to work correctly after the renaming. These heuristics aim to ensure that those perturbations are performed systematically, maintaining the database's integrity and compatibility with SQL queries. The details for each perturbation type are as follows:

\uline{Add columns}: we input both the table name and all of its column names and data types to GPT-3.5 and prompt it to generate multiple column names and their corresponding data types that are suitable and congenial with reason and common sense given the current scenario, and prompt GPT-3.5 don’t generate the column names that have the similar meaning with the existing input column names. Then we add a heuristic guarantee to filter out the redundant columns if the generated column names are repeated. These synthesized columns are then randomly inserted into the relevant tables. Notably, both the NLQ and the gold SQL remain unchanged during this process.

\uline{Remove columns}: We randomly eliminate
columns from the given schema, ensuring that the removed columns do not appear in the gold SQL query. Again, the NLQ and the gold SQL are kept fixed during this operation.

\uline{Remove columns in gold SQL}: In this scenario, we randomly remove columns from the schema, specifically targeting those referenced in the gold SQL query. As a result, the gold SQL becomes invalid. Instead, we use the response ``The given column information is insufficient to generate an SQL query to answer the question" as the ground truth.

\uline{Rename columns}: as Figure \ref{fig:data_collection} (a) shows, we input both the table name and all of its column names and data types to GPT-3.5. We randomly select multiple column names and their data types and prompt GPT-3.5 to generate similar, context-appropriate names. These synthesized names replace the original column names. In addition, in order to maintain the correctness of the relationship among the tables, If the column in one table has been renamed, we will also rename the foreign keys in other tables if those columns reference the renamed one. We also revise gold SQL accordingly to ensure that the revised schema and gold SQL remain aligned with the unchanged NLQ.

\textcolor{reviewer2}{
\uline{Split columns}: Since columns such as name, address, and date are often stored in more fine-grained formats in real-world databases (e.g., a full name split into first and last name; a date split into year, month and day; an address split into state, city and street, etc), we identify examples in BIRD dev set that involve these attributes and manually split the corresponding columns into finer-grained columns for evaluation. As these changes affect the structure of the gold SQL queries, we manually revise the gold SQL to reflect the updated schema. For the training set, we similarly select examples in BIRD train set involving name, address, or date, and use Claude 3.5 to synthesize the corresponding fine-grained columns and update the gold SQL accordingly.
}

\textcolor{reviewer2}{
% \uline{Merge columns}: As the reverse of column splitting, fine-grained columns are often stored as a single, more abstract column in real-world databases (e.g., first name and last name as full name; year, month, and day as date; state, city, and street as address). We identify relevant examples in the BIRD dev set and manually merge the corresponding fine-grained columns into their abstract forms for evaluation. These schema changes affect the gold SQL queries, which we revise accordingly. For training, we select similar examples from the BIRD train set and use Claude 3.5 to synthesize the merged schema and update the gold SQL.
\uline{Merge columns}:As the reverse of column splitting, we simulate more abstract column representations commonly seen in real-world databases (e.g., combining first and last name into full name; year, month, and day into date; state, city, and street as address). We identify relevant examples in the BIRD dev set and manually merge fine-grained columns, updating the gold SQL accordingly. For training, we apply the same strategy to the BIRD train set and use Claude 3.5 to synthesize the merged schema and update the gold SQL.}

\uline{Add tables}: We randomly add irrelevant tables to each question, and these tables are still in the same database as the relevant tables in BIRD. The original BIRD datasets guarantee that no different tables in their database can lead to alternative correct SQL answers. The tables added don't affect the NLQ and the gold SQL.

\uline{Remove tables}: In this scenario, we randomly remove tables from the relevant schema, which are referenced in the gold SQL query. As a result, the gold
SQL becomes invalid. Instead, we use the response “The given table information is insufficient to generate an SQL query to answer the question" as the ground truth.

\uline{Rename tables}: we input both the table name and all of its column names and data types to GPT-3.5. We randomly select one or multiple table names and prompt GPT-3.5 to generate similar, context-appropriate names. These synthesized names replace the original table names. In addition, in order to maintain the correctness of the relationship among the tables, we will also rename the foreign keys in other tables if they reference the renamed table. Finally, the table names in the gold SQL will also be renamed.

\uline{Split tables}:  as Figure \ref{fig:data_collection} (b) shows, we input both the table name and all of its column names and data types to GPT-4. We prompt GPT-4 to identify tables that can be logically divided into two or more smaller tables. Using GPT-4, we generate new table names and distribute the columns of the original table among the new tables in a contextually appropriate manner. The primary key in the original table will be copied into all the new tables after splitting. The gold SQL is revised by GPT-4 to reference the newly created tables, ensuring consistency across all components. We also manually check the new gold SQL to make sure it’s correct.

\uline{Merge Tables}: We select two or more related tables and combine them into a single table. GPT-4 is used to generate a suitable name for the merged table, and the columns from the original tables are consolidated under this new table. More concretely, the GPT4 is prompted to 1) copy all the primary key columns of the original tables to the new tables after merging, but only keep one of them as the primary key of the new table, and make others as the regular columns. 2) if the primary key columns in these two original tables are the same, then just keep one in the new table after merging. 3) when merging tables, if there are two columns not the primary key column but with the same names in the original tables, revise their column names accordingly to make them different when merging them into the new table. Finally, the gold SQL is updated by GPT-4 accordingly. We also manually check the new gold SQL to make sure it’s correct.

\noindent\textbf{Quality Control:} \nop{To ensure the data quality in \texttt{EvoSchema}, we request five annotators with SQL expertise to carefully review the splitting and merging tables cases, and manually correct the new gold SQLs if they are not correct.}
To ensure high-quality data in \texttt{EvoSchema}, we leverage advanced language models and rigorous human validation. Specifically, we use GPT-3.5 to generate synthesized column and table names and data types (only for columns) when adding or renaming are required. We randomly choose 200 generated examples to do manual review and reveal that GPT-3.5 demonstrates a strong understanding of the input context, effectively generating names that meet our requirement. For more complex operations, such as splitting or merging tables, we utilize the capabilities of more powerful GPT-4 to handle both schema changes and corresponding SQL modifications with high accuracy.

To complement these automated processes, we engaged five annotators with substantial SQL expertise to carefully review cases involving complex schema transformations. Annotators validated and, where necessary, manually corrected the generated gold SQL queries to ensure correctness and alignment with the modified schemas. To further enhance reliability, we implemented cross-validation by assigning complex cases to multiple annotators and resolving discrepancies through discussion or consensus. This combination of advanced AI tools and meticulous human review ensures that \texttt{EvoSchema} maintains a robust and accurate benchmark, faithfully reflecting real-world schema evolution scenarios.

\noindent\textcolor{reviewer3}{\textbf{Cost Analysis:}\label{cost_analysis}
We have 1.5K split-table examples and 1.1K merge-table examples requiring human verification. Among the split examples, 1.1K are relatively simple and take approximately 3 minutes each to verify, while the remaining 0.4K are more complex and require about 7 minutes each—totaling roughly 100 hours. For the merge-table examples, 0.8K are simple (3 minutes each) and 0.3K are complex (7 minutes each), amounting to approximately 75 hours. \textcolor{reviewer_shepherd}{Note this manual effort was for curating the evaluation data, not the training data.\nop{As with any benchmark, this one-time effort ensures the accuracy and reliability of evaluation results, allowing the broader community to use the data without repeating the verification process. In contrast,} Our training data is generated entirely automatically without any human annotation or manual verification.} Our analysis also indicates that LLM-generated split and merge tables include around 30\% low-quality data, underscoring the need for careful human validation for these two types.} 

\subsection{Comparison with Existing Benchmarks}
\label{comparison_with_existing_benchmarks}

\texttt{EvoSchema}, as presented in Table \ref{tab:comparison_with_existing_benchmarks}, introduces a comprehensive and unique taxonomy for evaluating models' behavior under the impact of schema evolution on SQL queries, distinguishing itself from other benchmarks like Dr.Spider \citep{chang2023drspider}, ADVETA \citep{pi-etal-2022-towards}, MT-TEQL \citep{mt-teql} and FootballDB \citep{footballdb}. Unlike Dr.Spider and ADVETA, which focus on limited perturbations such as column renaming and additions, \texttt{EvoSchema} encompasses a broader range of transformations, including adding, removing, renaming, splitting and merging at both the column level and table level. This diversity allows for testing systems under realistic and dynamic schema evolution scenarios. Furthermore, while MT-TEQL includes a variety of perturbations, it only modifies the columns not mentioned in the SQL which does not consider the impact of schema evolution on SQL directly. \texttt{EvoSchema} uniquely integrates schema evolution with its effects on SQL queries, enabling evaluation of models in environments that closely mimic real-world database evolution challenges. \textcolor{reviewer1}{Different from FootballDB \citep{footballdb} which mainly restructures schema to reduce foreign key mappings among tables and reduce JOIN paths for SQL, we define a more configurable, systematical and structured schema evolution taxonomy. Besides, our provided schema evolution and synthesis framework allows us to explore the schema change on multiple databases easily, while FoodballDB is only limited to one database.} Finally, for the seed data selection, compared to Spider, which is commonly used to study
text-to-SQL robustness, BIRD features more intricate, realistic, and
extensive databases, as well as more complex SQL queries that include keywords often missing in Spider. These distinctions make \texttt{EvoSchema} particularly well-suited for studying how systems adapt to evolving schemas, advancing beyond the simpler or less holistic setups of prior benchmarks.

\subsection{Data Statistics}
\begin{table}[ht]
\caption{Data statistics of \texttt{EvoSchema}. ``Original'' refers to the original BIRD dataset; ``Column Manipulation'' refers to applying the column-level operations on the columns of the original BIRD data; ``Table Manipulation'' refers to applying the table-level operations on the tables of the original BIRD data. ``*'': the evaluation data for calculating execution accuracy. We synthesize values to reconstruct the database after schema evolution, and filter out those not executable by gold SQL, which results in the smaller size of the evaluation data for calculating execution accuracy.}

\small
\resizebox{1.0\linewidth}{!}{
\centering
\begin{tabular}{@{}l|ccc|cccc|cccc@{}}
\toprule
\multicolumn{12}{c}{Data Statistics} \\
\midrule
 \multirow{2}{*}{Perturbation Type}  &\multirow{2}{*}{Train} & \multirow{2}{*}{Eval} & \multirow{2}{*}{Eval*} & \multicolumn{4}{c}{Manipulated Items/Table}  & \multicolumn{4}{c}{Manipulated Items/Query}  \\
 &&& &Min & Mean & Median & Max &Min & Mean & Median & Max  \\
 \midrule
 Original & 9426 & 1534 & 1068 &-&-&-&-&-&-&-&-  \\
 \midrule
&&& & \multicolumn{8}{c}{Column Manipulation} \\
\midrule
Add Columns  & 9219 & 1506 & 846 & 1 & 5.7 & 3 & 83 & 1 & 5.9 & 4 & 43\\
Remove Columns &  9426 & 1534 & 1076 & 1 & 6.2 & 2 & 87 & 1 & 6.9 & 3 & 70 \\ 
Remove Col in SQL & 9424 & 1534 & - & 1 & 2.5 & 2 & 8 & 1 & 2.5 & 2.5 & 6 \\
Rename Columns & 9385 & 1533 & 947 &  1 & 4.3 & 3 & 46 & 1 & 4.4 & 3 & 46 \\
\textcolor{reviewer2}{Split Columns} & 140 & 37 & 37 & 1 & 2 & 2 & 4 & 1 & 2 & 2 & 4\\
\textcolor{reviewer2}{Merge Columns} & 148 & 44 & 44 & 2 & 3 & 3 & 4 & 2 & 3 & 3 & 4\\
 \midrule
  &&& & \multicolumn{8}{c}{Table Manipulation} \\
\midrule
Add Tables &  9387 & 1530 & 1014  & - & - & - & - & 1 & 2 & 2 & 3    \\ 
Remove Tables & 7212 & 1171 & -  & - & - & - & - & 1 & 1 & 1 & 1 \\
Rename Tables &  9392 & 1534 & 1063 & - & - & - & - & 1 & 1.5 & 1 & 4  \\ 
Split Tables & 9254 & 1515 & 824 & - & - & - & - & 1 & 2.6 & 3 & 5 \\
Merge Tables & 6930 & 1139 & 569 & - & - & - & - & 2 & 2 & 2 & 2\\

\bottomrule
\end{tabular}
}
\label{tab:data_stat_plus_manipulated_stat}
\end{table}

Table \ref{tab:data_stat_plus_manipulated_stat} provides an overview of the data statistics in \texttt{EvoSchema}, showcasing the various perturbation types applied to the original BIRD dataset.  ``Column Manipulation" refers to applying
the column-level operations on the columns of the original BIRD data; ``Table Manipulation" refers to applying the table-level
operations on the tables of the original BIRD data. All the perturbed data are obtained by applying column manipulation or table manipulation on the original BIRD dataset. ``Manipulated Items" shows the size of the altered columns or the tables. ``Manipulated Items/Query" refers to the number of columns or tables modified in the schema for each SQL query, specifically targeting the tables relevant to generating that query. For ``Split Tables," ``Manipulated Items/Query" represents the number of tables each original table is split into. For ``Merge Tables", ``Manipulated Items/Query" indicates the number of tables combined into a single table.

\section{Training Paradigm} 

In our work, we propose a new training paradigm to enhance the model's robustness against different schema evolution. For each \textit{\textless NLQ, relevant schema, SQL\textgreater} triple, we fix the NLQ in the training data, and augment each triple with different schema designs, which may or may not lead to SQL change. Consequently, we obtain multiple triples that can be derived from each of the original triples. We train the model by learning multiple schema designs and SQLs to the original question mappings, which can improve the model's ability to identify the correct relationships among different tables and columns to the question, and can better distinguish the difference among different schema designs. Through this procedure, the model can avoid learning spurious patterns better and therefore enhance the robustness against different schema evolution types.

\section{Experiment Setup}
\nop{In the following sections, we detail our experimental setup, including the specific schema perturbation scenarios tested, the models trained, and the evaluation metrics used. We then present the results of our experiments, demonstrating how \texttt{EvoSchema} improves model robustness across various schema evolution scenarios.} \nop{and discuss the implications of these findings for future research and applications.}

\nop{\subsection{Dataset}
For our experiments, we utilize the BIRD \citep{bird} dataset, which is specifically designed for the text-to-SQL task. Compared to Spider \citep{yu-etal-2018-spider}, which is commonly used to study text-to-SQL robustness, BIRD features more intricate, realistic, and extensive databases, as well as more complex SQL queries that include keywords often missing in Spider. BIRD consists of NLQs, corresponding database schemas, and gold SQL queries and encompasses a wide range of real-world database scenarios, which provides a robust foundation for evaluating the performance of models in translating NLQs into SQLs.

Schema Perturbations: To evaluate the robustness of the text-to-SQL models, we use the BIRD dataset not only in their original form but also augmented with various column-level and table-level schema perturbations. We ensure that the NLQs remain fixed, while the schema and SQL queries are adjusted as necessary to reflect the changes introduced by our perturbations. We follow the standard train/dev split provided with BIRD, and apply all the perturbations on both training data and evaluation data. The data statistics are in Table \ref{tab:data_stat_plus_manipulated_stat} and the examples of different perturbation types are in Figure \ref{fig:appendix_examples}. \nop{\hs{did you mention the statistics of the perturbed/all data somewhere?}}
}

% Motivation for Using BIRD: BIRD is chosen due to its rich variety of schema structures and natural language questions, making it a suitable benchmark for testing models' abilities to handle schema variations and generate accurate SQL queries. The diversity of the BIRD dataset allows us to evaluate model performance across a broad spectrum of schema complexities and perturbation scenarios.

% Q: where to put the input format? DDL (maybe directly show a table to list examples for all the types)

\begin{table*}[t]
\caption{Evaluation on \texttt{EvoSchema}. ``w/'': the model is trained by merging the original data and all the perturbation training types together; ``w/o'': the model is only trained on the original training data. The \textbf{best performance} for each model is in bold, and \textcolor{mydarked}{\textbf{red}} shows a larger gain. ``-'': some of the relevant tables are removed so there should be no gold SQL used to calculate the metrics here.}
% \vspace{-0.5\baselineskip}
\label{tab:bird_main_results2}
\small
\centering
\resizebox{0.7\linewidth}{!}{
\begin{tabular}{@{}l|cc|cc|cc|cc|cc@{}}
\toprule
 \multirow{2}{*}{Perturbation Type} 
 & \multicolumn{2}{c}{Code Llama} & \multicolumn{2}{c}{Mistral} & \multicolumn{2}{c}{Llama 3} & \multicolumn{2}{c}{SQLCoder} & GPT-3.5 & GPT-4 \\
 & w/o & w/ & w/o & w/ & w/o & w/ & w/o & w/  \\
 \midrule
 \multicolumn{11}{c}{Table Match F1} \\
 \midrule
Original & 89.77 & \textbf{90.42} & 89.58 & \textbf{90.62} & \textbf{89.96} & 89.53 & 89.69 & \textbf{90.64} & 87.28 & 88.98 \\
\midrule
Add Columns & 89.73 & \textbf{90.27} & 89.65 & \textbf{90.03} & 89.08 & \textbf{89.70} & 89.30 & \textbf{90.52} & 86.35 & 88.12  \\
Remove Columns & 89.82 & \textbf{90.24} & 89.89 & \textbf{90.66} & \textbf{90.09} & 89.82 & 89.81 & \textbf{90.54} & 87.18 & 88.87 \\ 
Rename Columns & \textbf{85.28} & 85.07 & \textbf{84.32} & 84.27 & \textbf{83.74} & 82.92 & \textbf{85.32} & 84.93 & 81.73 & 83.20  \\ 
\textcolor{reviewer2}{Split Columns} & 83.78 & \textcolor{mydarked}{\textbf{89.19}} & 83.78 & \textcolor{mydarked}{\textbf{88.29}} & 81.08 & \textcolor{mydarked}{\textbf{85.14}} & 86.49 & \textcolor{mydarked}{\textbf{88.29}} & 81.44 & 86.31 \\
\textcolor{reviewer2}{Merge Columns} & \textbf{88.65} & 87.23 & 87.23 & \textbf{89.72} & \textbf{88.65} & 86.17 & 87.23 & 87.23 & 83.17 & 89.36\\
\midrule
Add Tables & 57.88 & \textcolor{mydarked}{\textbf{89.50}} & 57.67 & \textcolor{mydarked}{\textbf{89.30}} & 55.11 & \textcolor{mydarked}{\textbf{88.51}} & 57.44 & \textcolor{mydarked}{\textbf{89.38}} & 83.54 & 85.79 \\ 
Remove Tables & - & - & - & - & - & - & - & - & - & - \\
Rename Tables & 88.84 & \textbf{90.32} & 89.40 & \textbf{90.56} & 87.18 & \textbf{89.14} & 89.40 & \textbf{90.48} & 87.02 & 88.45  \\ 
Split Tables & 71.99 & \textcolor{mydarked}{\textbf{81.55}} & 66.12 & \textcolor{mydarked}{\textbf{80.87}} & 71.08 & \textcolor{mydarked}{\textbf{80.12}} & 72.52 & \textcolor{mydarked}{\textbf{81.92}} & 77.52 & 80.68  \\
Merge Tables & 85.29 & \textbf{87.03} & 83.39 & \textcolor{mydarked}{\textbf{86.91}} & 81.68 & \textcolor{mydarked}{\textbf{86.48}} & 84.80 & \textbf{86.35} & 83.04 & 86.99 \\
\midrule
\textcolor{reviewer3}{MacroAvg} & 83.10 & \textcolor{mydarked}{\textbf{88.08}} & 82.10 & \textcolor{mydarked}{\textbf{88.12}} & 81.77 & \textcolor{mydarked}{\textbf{86.75}} & 83.20 & \textcolor{mydarked}{\textbf{88.03}} & 83.83 & 86.68\\
\midrule
\multicolumn{11}{c}{Column Match F1} \\
\midrule
 Original & 80.66 & \textbf{81.64} & 81.10 & \textbf{82.36} & \textbf{79.13} & 78.72 & 81.52 & \textbf{81.97} & 78.28 & 80.78\\
\midrule
Add Columns & 78.26 & \textbf{80.27} & 79.16 & \textbf{80.18} & 75.79 & \textbf{76.87} & 79.09 & \textbf{80.46} & 75.03 & 78.58 \\
Remove Columns & 82.67 & \textbf{82.75} & 83.09 & \textbf{84.00} & \textbf{81.56} & 80.69 & \textbf{83.20} & 83.18 & 80.33 & 82.55 \\
Rename Columns & 76.50 & \textbf{76.94} & 76.35 & \textbf{76.73} & \textbf{72.24} & 71.07 & 76.84 & \textbf{77.38} & 73.40 & 75.90 \\
\textcolor{reviewer2}{Split Columns} & 71.22 & \textcolor{mydarked}{\textbf{81.81}} & 70.24 & \textcolor{mydarked}{\textbf{80.41}} & 67.29 & \textcolor{mydarked}{\textbf{75.04}} & 74.50 & \textcolor{mydarked}{\textbf{79.92}} & 73.59 & 77.92\\
\textcolor{reviewer2}{Merge Columns} & 83.19 & \textbf{83.30} & 82.75 & \textbf{83.41} & 82.72 & \textbf{83.68} & 82.64 & \textbf{83.31} & 78.13 & 88.56\\
\midrule
Add Tables & 63.81 & \textcolor{mydarked}{\textbf{81.14}} & 65.39 & \textcolor{mydarked}{\textbf{81.09}} & 59.36 & \textcolor{mydarked}{\textbf{77.96}} & 62.91 & \textcolor{mydarked}{\textbf{81.23}} & 76.45 & 79.32 \\
Remove Tables  & - & - & - & - & - & - & - & -  & - & - \\
Rename Tables & 79.60 & \textbf{80.91} & 80.32 & \textbf{81.29} & \textbf{77.49} & 77.46 & 80.77 & \textbf{81.79} & 77.78 & 80.04 \\
Split Tables & 75.30 & \textcolor{mydarked}{\textbf{78.45}} & 73.87 & \textcolor{mydarked}{\textbf{78.11}} & 73.81 & \textbf{73.95} & 75.83 & \textcolor{mydarked}{\textbf{78.59}} & 74.89 & 77.41 \\
Merge Tables & 65.56 & \textbf{67.09} & 64.12 & \textcolor{mydarked}{\textbf{
67.46}} & 63.50 & \textbf{64.40} & 65.57 & \textbf{67.29} & 63.23 & 68.13 \\
\midrule
\textcolor{reviewer3}{MacroAvg} & 75.68 & \textcolor{mydarked}{\textbf{79.43}} & 75.64 & \textcolor{mydarked}{\textbf{79.50}} & 73.29 & \textcolor{mydarked}{\textbf{75.98}} & 76.29 & \textcolor{mydarked}{\textbf{79.51}} & 75.11 & 78.92\\

\bottomrule
\end{tabular}
}
\end{table*}

\subsection{Training and Evaluation Settings}
\noindent\textbf{Training Setting:} We choose four open-source models: Code Llama-7B \citep{rozière2024codellamaopenfoundation}, Mistral-7B \citep{jiang2023mistral7b}, Llama 3-8B \citep{dubey2024llama3herdmodels} and SQLCoder-7B \footnote{https://huggingface.co/defog/sqlcoder-7b-2\label{myfootnote}} and two closed-source models: GPT-3.5 \footnote{https://openai.com/chatgpt/} and GPT-4 \citep{openai2024gpt4technicalreport} for our experiments. For these four open-source models, we explore two settings:
1) without perturbation types: the model is trained on the original training data without any perturbation types introduced during training. 2) with perturbation types: the model is trained by merging both the original training data and the perturbation training data. For closed-source models, we only use them for evaluation.

\textcolor{reviewer1}{\noindent\textbf{Evaluation Setting:}
For all the closed-source models and the finetuned open-sourced models, we evaluate them under two settings: 1) without perturbation types: this setting uses the standard, unaltered original evaluation data to evaluate the model performance. 2) with perturbation types: the models are evaluated on data where different perturbations are introduced. By comparing the model performance under these two settings, we can assess how resilient the finetuned models and GPT models are to schema evolution in NL2SQL.
This setup provides a comprehensive evaluation of model performance in both standard and perturbed environments, allowing for detailed analysis of robustness and adaptability across different models and schema evolution types.}

% \subsection{different ablation study settings}

\subsection{Evaluation Metrics}
1) Table Match F1: this score is a metric to measure how well the model correctly identifies the relevant tables required to generate a valid SQL query. The F1 score is a harmonic mean of precision and recall, where the precision is the percentage of tables correctly predicted out of all tables predicted by the model and the recall is the percentage of tables correctly predicted out of all the actual tables that should have been selected. The Table Match F1 score combines these two metrics to provide a balanced evaluation, which can assess the ability of text-to-SQL models to correctly identify the required tables from the database schema to form accurate queries. A higher Table Match F1 indicates better performance in selecting the correct tables for the SQL query.

2) Column Match F1: this score is to evaluate how accurately the model identifies the relevant columns required to generate a valid SQL query from a natural language input. Like the Table Match F1, it measures the balance between precision and recall but is applied specifically to the columns of the database. A higher Column Match F1 score indicates better performance in selecting the right columns for the SQL query.

3) Execution Accuracy: this metric measures whether the predicted SQL query can return the correct results as the gold SQL when executing against a database. \nop{Since the schema evolution may lead to database restructure and there are no existing values for the new database after schema change, we synthesize values to create new databases and execute the new gold SQLs after schema evolution on them. Due to the complexity of the value synthesis and huge manual efforts to ensure an executable database for each instance, we filter out the cases where synthesized database is not executable by new gold SQL. This procedure can lead to a relatively smaller size of the evaluation data compared with table match F1 and column match F1. }\nop{so we mainly use the other two metrics as the main metrics.} \nop{\hs{do you also compute F1 on the returned records here?}}

\subsection{Training and Evaluation Details}
We choose Code Llama-7B \citep{rozière2024codellamaopenfoundation}, Mistral-7B \citep{jiang2023mistral7b}, Llama 3-8B \citep{dubey2024llama3herdmodels} and SQLCoder-7B \footref{myfootnote} as our open-source base models. We fine-tune these models with Huggingface
transformers library \citep{wolf2020huggingfacestransformersstateoftheartnatural}. For the perturbation training, we merge
all the perturbation data and randomly shuffle them as our final training data.
We use a learning rate of 2e-5 for training Code Llama, Llama 3 and SQLCoder, and 5e-6 for training Mistral. Our batch size is 4. We train all the models on 4 A100
80GB GPUs and use a cosine scheduler with a 0.03
warm-up period for 6 epochs. We employ FSDP \citep{zhao2023pytorchfsdpexperiencesscaling} to efficiently train
the model. We set the max input length of training as 1024 and the max output length of inference as 500. For inference, we use vllm \citep{wolf2020huggingfacestransformersstateoftheartnatural} for batch evaluation, and we set the batch size as 16. We do the inference on an 80G A100 GPU. For closed-source LLMs, we use Azure OpenAI API\footnote{https://learn.microsoft.com/en-us/azure/ai-services/openai/reference}. We use the 2023-12-01-preview version for GPT-4, and 2023-07-01-preview version for GPT-3.5.

\subsection{Baselines} 
\label{baselines}
We add in-context learning \citep{gao2024text} and more advanced method: CHESS \citep{chess_sql} as the baselines for comprehensive comparison. \textcolor{reviewer2}{In order to test whether the in-context learning can help address the schema evolution issue, we randomly select three examples (each example is an \textit{\textless NLQ, database schema after evolution, gold SQL after schema evolution\textgreater} triple) as the demonstration in the prompt to help the models understand the schema after evolution (Table \ref{tab:bird_human_eval}).} \textcolor{reviewer3}{We also include CHESS, an advanced method for NL2SQL as a baseline. We apply the schema selection (SS) and candidate generation (CG) components developed in their work. For schema selection, we use advanced gpt-4o model to prune the database schema and remove the irrelevant tables and the irrelevant columns in the selected tables, ensuring only the most relevant tables and columns are passed into the model for SQL generation. To ensure a fair comparison with our primary fine-tuning approach, we use a fine-tuned Code Llama model trained without any schema perturbation data as the SQL generation model. This setup allows us to isolate and evaluate the effectiveness of a schema selection and pruning component in addressing schema evolution. The results are shown in Table \ref{tab:bird_human_eval}. \nop{Our hypothesis is that a strong schema selection mechanism can mitigate the impact of schema changes while preserving the essential information upfront, thereby reducing both the burden and the noise in downstream SQL generation. }}
\section{Results and Analysis}

\subsection{Main Results}
\label{main_results}

\begin{table}[hb]
  \centering
\caption{\textcolor{reviewer_shepherd}{Human Evaluation on \texttt{EvoSchema}. ``ZS'' refers to zero-shot, which prompts models without any examples. ``ICL'' refers to in-context learning, which prompts models with three demonstration examples. ``w/o'' means fine-tuning model without perturbation training data; ``w/'' means fine-tuning model with perturbation training data. Bold color indicates the best performance among each row.}}
% \vspace{-0.5\baselineskip}
\label{tab:bird_human_eval}
\small
\resizebox{0.8\linewidth}{!}{
\begin{tabular}{@{}l|ccccc@{}}
\toprule
\multicolumn{6}{c}{Human Evaluation on \texttt{EvoSchema}} \\
\midrule
 \multirow{2}{*}{Perturbation Type} & \multicolumn{2}{c}{GPT-4} & \multicolumn{2}{c}{Code Llama} & CHESS$_{SS+CG}$\\
 & ZS & ICL & w/o & w/\\
 \midrule

Original & 62 & 58 & \textbf{65} & 64 & 63\\
\midrule
Add Columns & 59 & 55 & 62 & 61 & \textbf{66}\\
Remove Columns & 65 & 61 & \textbf{66} & 63 & 64 \\ 
Rename Columns & 57 & 56 & 57 & 57 & \textbf{62} \\ 
Split Columns & 46 & 59 & 41 & \textbf{62} & 49 \\
Merge Columns & 68 & 66 & 70 & \textbf{70} & 66 \\
\midrule
Add Tables & 56 & 55 & 46 & \textbf{62} & 57\\ 
Remove Tables & - & - & - & - & -\\
Rename Tables & 58 & 60 & \textbf{64} & 61 & 61 \\ 
Split Tables &  57 & 53 & 48 & \textbf{60} & 53 \\
Merge Tables &  55 & 57 & 54 & \textbf{58} & 53\\
\midrule
MacroAvg & 58 & 58 & 57 & \textbf{62} & 59 \\

 \bottomrule
\end{tabular}
}
\end{table}
\begin{table}[hb]
  \centering
\caption{\textcolor{reviewer3}{Execution Accuracy on \texttt{EvoSchema}. ``w/'': the model is trained with all the perturbation types; ``w/o'': the model is only trained on the original training data.}}
% \vspace{-0.5\baselineskip}
\label{tab:bird_exec_acc}
\small
\resizebox{1.0\linewidth}{!}{
\begin{tabular}{@{}l|cc|cc|cc|cc|cc@{}}
\toprule
\multicolumn{11}{c}{\textcolor{reviewer3}{Exec Acc on \texttt{EvoSchema}}\nop{\tablefootnote{Since the schema evolution may lead to database restructure and there will be no existing values for the new database after schema change. We synthesize values to create new databases after schema evolution and execute the new gold SQLs after schema evolution on the new databases.}}} \\
\midrule
 \multirow{2}{*}{Perturbation Type} & \multicolumn{2}{c}{Code Llama} & \multicolumn{2}{c}{Mistral} & \multicolumn{2}{c}{Llama 3} & \multicolumn{2}{c}{SQLCoder} & GPT-3.5 & GPT-4 \\
 & w/o & w/ & w/o & w/ & w/o & w/ & w/o & w/ \\
 \midrule
 Original & 58 & 57 & 59 & 58 & 55 & 51 & 58 & 58 & 44 & 47  \\
\midrule
Add Columns & 57 & 55 & 56 & 56 & 52 & 49 & 55 & 57 & 43 & 46\\
Remove Columns & 59 & 57 & 60 & 58 & 56 & 53 & 60 & 58 & 45 & 47 \\ 
Rename Columns & 54 & 52 & 55 & 54 & 49 & 47 & 56 & 55 & 43 & 45 \\ 
\textcolor{reviewer2}{Split Columns} & 41 & \textcolor{mydarked}{\textbf{62}} & 35 & \textcolor{mydarked}{\textbf{54}} & 38 & \textcolor{mydarked}{\textbf{49}} & 43 & \textcolor{mydarked}{\textbf{67}} & 41 & 46\\
\textcolor{reviewer2}{Merge Columns} & 70 & 70 & 70 & 70 & 73 & 73 & 66 & \textcolor{mydarked}{\textbf{82}} & 61 & 68\\
\midrule
Add Tables & 40 & \textcolor{mydarked}{\textbf{58}} & 39 & \textcolor{mydarked}{\textbf{58}} & 37 & \textcolor{mydarked}{\textbf{52}} & 40 & \textcolor{mydarked}{\textbf{57}} & 44 & 48\\ 
Remove Tables & - & - & - & - & - & - & - & - & - & -\\
Rename Tables & 56 & 55 & 55 & 56 & 52 & 50 & 56 & 55 & 43 & 47 \\ 
Split Tables & 38 & \textcolor{mydarked}{\textbf{46}} & 36 & \textcolor{mydarked}{\textbf{48}} & 40 & 41 & 43 & \textcolor{mydarked}{\textbf{49}} & 40 & 47  \\
Merge Tables & 43 & 45 & 45 & 46 & 42 & 44 & 47 & 46 & 37 & 45\\
\midrule
\textcolor{reviewer3}{MacroAvg} & 52 & \textcolor{mydarked}{\textbf{56}} & 51 & \textcolor{mydarked}{\textbf{56}} & 49 & \textcolor{mydarked}{\textbf{51}} & 52 & \textcolor{mydarked}{\textbf{58}} & 44 & 49\\

\bottomrule
\end{tabular}
}
\end{table}

As \textcolor{reviewer3}{Table \ref{tab:bird_main_results2} and Table \ref{tab:bird_exec_acc}} show, we train Codellama, Mistral, Llama3 and SQLCoder on the original BIRD training data with and without different perturbation types, and evaluate the model on the original BIRD evaluation data and different perturbation types. We observe:
\nop{Adding the perturbation data during training: 1) does not sacrifice the performance of the original evaluation data; 2) achieves comparable or better results on different perturbation types. Column-level schema changes are relatively minor compared with table-level schema changes. We can see the models perform better on both the column-level and table-level perturbation types in general, which shows the models are robust to both minor schema evolution and major schema evolution.}

\textcolor{reviewer3}{\noindent\textbf{The models trained on different perturbation types are more robust to the schema variation on average, and demonstrate high robustness on the table-level schema evolution.} While adding the perturbation data during training leads to a slight Exec Acc (EX) drop for original non-evolved evaluation data, adding, removing and renaming column types, it achieves significantly better results on splitting columns and table-level perturbation types. By comparing these four models' performance with and without the perturbation data, we observe that for splitting columns, the model trained with perturbation data can achieve up to 5.4 points gain for table match F1, 10.6 points gain column match F1 and 24 points gain for EX; for adding tables, the model trained with perturbation data can achieve up to 33 points gain for table match F1, 18 points gain for column match F1 and 19 points for EX; for splitting tables, the model trained with perturbation data can achieve up to 14 points gain for table match F1, 4.2 points gain for column match F1 and 12 points for EX; for merging tables, the model trained on perturbation data can achieve up to 4.8 points gain on table match F1 and 3 points gain for column match F1. We hypothesize that this is because the perturbation augmented data is particularly beneficial for handling substantial schema changes, but may introduce minor
noise in simpler schema changes where the model trained without perturbation data has already maximally learned the patterns.} \textcolor{reviewer_shepherd}{To better understand the slight performance gap under simpler column-level perturbations, we conducted error analysis and case studies to compare models trained with and without perturbed data. We observed two types of errors that lead to this phenomenon: (1) Spurious or missing conditions in the WHERE clause. For instance, given the question "What is the element with the atom ID of TR004\_7 in molecule that is not carcinogenic?", the model trained with perturbation (``w/") misses the condition T2.label = '-' in WHERE clause, while the ``w/o" model includes it correctly. However, in another case, 'How many transactions were paid in CZK on the morning of 2012/8/26?', the ``w/" model introduces an unnecessary WHERE condition: T1.TransactionID BETWEEN 1 AND 1000, which is not part of the gold SQL. (2) Incorrect column selection in SELECT or WHERE clauses. For example, for the question "Among the patients followed at the outpatient clinic, how many of them have a normal level of alkaliphophatase?", the ``w/" model predicts T1.Description instead of T1.Admission in WHERE clause, while the ``w/o" model selects the correct column. Similarly, in the question "Which group does superhero A-Bomb belong to?", the ``w/" model selects T2.team\_affiliation instead of the correct T2.race. These examples suggest that while training with perturbed data can improve general robustness, especially beneficial for handling substantial schema changes, it may also introduce minor noise that misleads in condition or column selection under simpler perturbations.}

\textcolor{reviewer3}{\noindent \textbf{Closed-source models are robust to different scheme evolution types in general.} As table \ref{tab:bird_main_results2} and \ref{tab:bird_exec_acc} show, we compare the model performance on GPT models and four open-source models trained with and without perturbation types. We observe that:  the GPT models' performance are relatively stable across different perturbation types compared to the original non-evolved test set. In contrast, fine-tuned open-source models without perturbation training data exhibit significant performance
drops—particularly on split columns, add tables, split tables, and merge tables—which introduce larger
schema changes. We hypothesize that the stability and robustness of closed-source models stems from
broader pretraining exposure and stronger internal schema reasoning capabilities, while the open-source models trained without perturbation types are more sensitive due to limited training on diverse schema variations. This motivates the need to
fine-tune open-source models with perturbation training data to improve their generalization under schema
evolution. \textit{We notice that comparing the model performance on the open-source LLMs and closed-source LLMs, the models trained with perturbation data have better performance than GPT models on both column-level and table-level perturbation evaluation data.} This indicates that our models trained with perturbation data are more robust than GPT models.}

\noindent\textbf{Table-level perturbation has a larger impact than column-level perturbation on the model performance.} As Table \ref{tab:bird_main_results2} and \ref{tab:bird_exec_acc} show, comparing with the performance on the original evaluation data: adding tables and splitting tables will lead to a significant table match F1 drop; adding tables, splitting tables and merging tables will lead to a significant column match F1 drop. This phenomenon indicates that adding tables or splitting tables easily confuses the models in choosing the correct tables to generate the SQL query. For merging tables, even though the model can correctly choose tables, it's a bit hard for the model to pick up the correct columns when the columns from different tables go into the same table. While for the column-level performance, there are limited differences with the performance on the original data except for splitting columns.

\nop{\textit{Compare the model performance on column-level and table-level perturbation evaluation, the column match F1 is worse than table match F1 in general.} Since the column size is much larger than the table size, it's easier for the model to select the table correctly than columns,}

\noindent\textbf{Reducing table schema complexity is beneficial for model performance.} Compare the model performance on column-level perturbation evaluation and the original evaluation data, adding columns results in a decrease in column match F1, whereas removing columns leads to an increase in column match F1. It indicates simpler table schema is beneficial for models to select columns, as removing columns simplifies the table schema while adding columns makes the table schema more complex.

% \textit{Compare the model performance on table-level perturbation evaluation and the original evaluation, table-level perturbation can lead to both the table match F1 and column match F1 drop in general.} 

\vspace*{-0.5\baselineskip}
\subsection{Comparison of Different Baselines}
\label{baseline_comparison}
As \texttt{EvoSchema} has a large scale of the test set and we need to call GPT-4 and GPT-4o API for in-context learning and CHESS respectively, to save the cost, \textcolor{reviewer_shepherd}{we randomly select 200 examples for the raw BIRD test set and also from each perturbation type to compare different baselines.} We compare GPT-4 zero-shot prompting, GPT-4 3-shot in-context learning, CodeLLama trained with and without perturbation training data and CHESS (with schema selection (SS) and candidate generation (CG)) on our downsampled test set. Since we found that Exec Acc can still make mistakes when different SQL queries produce the same results sometimes even they don't align with the NLQ, or sometimes both the gold SQL and wrong predicted SQL return the empty which may mislead the evaluation, we use human evaluation here for more precise evaluation. \textcolor{reviewer2}{As Table \ref{tab:bird_human_eval} shows, compared to GPT-4 zero-shot (ZS), in-context learning (ICL) shows a significant advantage only on the split columns perturbation, while performing slightly better or worse on other types. This suggests that ICL is not consistently effective for handling schema evolution. We hypothesize this is because the demonstration examples in ICL cannot cover the full range of schema and SQL changes; thus, for examples that differ significantly from the demonstrations, ICL offers limited benefit. However, for split columns, where changes commonly involve patterns like name, address, or date splits, the demonstrations generalize better—making ICL more effective in this case.\nop{compared with GPT-4 zero-shot (ZS) results, in-context learning (ICL) only largely exceeds it on `split columns' type and for other perturbation types, it's slightly better or worse. This indicates that in-context learning is not always helpful for addressing the schema evolution problem. We hypothesize that it's because the demonstration examples in the ICL can not cover all the cases of the schema change and SQL change, so for those examples that are highly different with the demonstration examples, ICL can not bring benefit. But for `split columns' types, since the column changes always bind to several types such as the name, address and date split, so the demonstration examples make ICL work better on this perturbation type.}} \textcolor{reviewer3}{For CHESS, we use GPT-4o—a powerful closed-source model—for schema selection and pruning, and Code Llama without perturbation training (CodeLlama w/o) as the SQL generation model. CHESS achieves the best performance on add columns and rename columns, and significantly outperforms CodeLlama w/o on split columns, add tables, and on average. This highlights the importance of accurate schema selection and pruning in improving SQL generation. However, we also observe that errors at the pruning stage can propagate, leading to degraded performance. Specifically, in merge columns and merge tables cases, CHESS tends to over-prune, omitting relevant schema information and resulting in worse performance than CodeLlama w/o.\nop{For CHESS, we use GPT-4o, an advanced and powerful closed-source model for schema selection and pruning, and use the Code Llama trained without perturbation training data (CodeLlama w/o) as the SQL generation model. We found that CHESS achieves the best results on `add columns', `remove columns' and `rename columns' types among all the baselines, and is also significantly better than `CodeLlama w/o' baseline on `split columns', `add tables' and on average. This indicates that schema selection and pruning is important and useful in general, especially if it can prune the schema accurately. But it's also likely that sometimes, the errors in this stage can propagate into the SQL generation component and lead to mistakes. We observe that for the `merge columns' and `merge tables' types, CHESS can prune more tables and columns and make the relevant schema information missing for the SQL generation and thus lead to worse performance than `CodeLlama w/o'.} Finally, we found that fine-tuning CodeLlama with perturbation training data is still needed, since this method gets the best performance among all the baselines on average across all types of evaluation data, and performs significantly better than others on `split columns', `add tables', `split tables' and `merge tables' types.} \textcolor{reviewer_shepherd}{We applied McNemar’s Test \citep{mcnemar1947note} to measure the statistical significance of performance differences between our method and each baseline. We computed p-values using the \texttt{statsmodels} package, considering differences statistically significant when 
$p<0.05$, which indicates that the improvement is unlikely due to random chance. Using this test, we observed our method achieved statistically significant improvements over three key baselines: GPT-4 in-context learning, fine-tuning without perturbed data, and CHESS (all with $p<0.05$).}
\nop{\section{Ablation Study}}

\nop{add a summary paragraph}

\begin{table}[t]
\caption{Perturbation type ablation on \texttt{EvoSchema}. The base model is Code Llama. ``both": the model is trained with both column-level perturbation and table-level perturbation types; ``w/o table-p": the model is trained without table-level perturbation types; ``w/o column-p": the model is trained without column-level perturbation types.}
\vspace{-0.5\baselineskip}
\label{tab:perturbation_type_effect}
\small
\resizebox{1.0\linewidth}{!}{
\centering
\begin{tabular}{@{}l|ccc|ccc@{}}
\toprule
\multicolumn{7}{c}{Perturbation Type Ablation} \\
\midrule
 \multirow{2}{*}{Perturbation Type} & \multicolumn{3}{c}{Table Match F1} & \multicolumn{3}{c}{Column Match F1} \\ 
 & both & w/o table-p & w/o column-p &  both & w/o table-p & w/o column-p \\
 \midrule
Original & 90.73 & 90.80 \scriptsize (+0.07) & 90.04 \scriptsize(-0.69) & 81.09 & 82.15 \scriptsize(+1.06) & 80.49 \scriptsize(-0.60) \\
\midrule
Add Columns & 90.86 & 90.80 \scriptsize(-0.06) & 89.75 \scriptsize(-1.11) & 79.63 & 80.81 \scriptsize(+1.18) & 77.29 \scriptsize(-2.34)\\
Remove Columns & 90.72 & 90.83 \scriptsize(+0.11) & 90.48 \scriptsize(-0.24) & 83.28 & 83.85 \scriptsize(+0.57) & 82.61 \scriptsize(-0.67) \\ 
Rename Columns & 85.35 & 85.38 \scriptsize(+0.03) & 84.57 \scriptsize(-0.78) & 76.49 & 77.53 \scriptsize(+1.04) & 75.17 \scriptsize(-1.32) \\ 
\midrule
Add Tables & 88.95 & 58.94 \textcolor{mydarked}{\scriptsize(-30.01)} & 88.57 \scriptsize(-0.38) & 79.87 & 64.11 \textcolor{mydarked}{\scriptsize(-15.76)} & 79.33 \scriptsize(-0.54)  \\ 
Remove Tables & - & - & - & - & - & -\\
Rename Tables & 90.54 & 90.77 \scriptsize(+0.23) & 89.29 \scriptsize(-1.25) & 81.13 & 81.51 \scriptsize(+0.38) & 79.33 \scriptsize(-1.80)\\ 
Split Tables & 80.71 & 73.28 \textcolor{mydarked}{\scriptsize(-7.43)} & 79.05 \scriptsize(-1.66)& 77.41 & 75.95 \scriptsize(-1.46) & 76.30 \scriptsize(-1.11)\\
Merge Tables & 88.72 & 87.87 \scriptsize(-0.85) & 86.83 \scriptsize(-1.89) & 68.40 & 68.26 \scriptsize(-0.14) & 67.08 \scriptsize(-1.32)\\

\bottomrule
\end{tabular}
}
\end{table}

\begin{table}[t]
\caption{Out of Scope Effect on \texttt{EvoSchema}. The base model is Code Llama. ``w/o": the model is trained without perturbation types; ``w/": the model is trained on the original data and all the perturbation types; ``+ OOS": the model is trained on the original data, perturbation types and two out-of-scope (OOS) perturbation types; ``+ OOS FP": The model trained with two OOS perturbation types makes an incorrect prediction on the original data and in-scope perturbation data; ``+ OOS TP": The model trained with two OOS perturbation types makes the correct prediction on the two OOS perturbation data; ``Tab": the model refuses to predict SQL due to the lack of table information; ``Col": the model refuses to predict SQL due to the lack of column information.}
% \vspace{-0.5\baselineskip}
\label{tab:out_of_scope_effect}
\small
\resizebox{1.0\linewidth}{!}{
\centering
\begin{tabular}{@{}l|ccc|ccc|c|c|c|c@{}}
\toprule
\multicolumn{11}{c}{Out of Scope Effect} \\
\midrule
 \multirow{2}{*}{Perturbation Type} & \multicolumn{3}{c}{Table Match F1} & \multicolumn{3}{c}{Column Match F1} & \multicolumn{2}{c}{+ OOS FP} & \multicolumn{2}{c}{ + OOS TP}\\ 
 & w/o & w/ & + OOS & w/o & w/ & + OOS & Tab & Col & Tab & Col\\
 \midrule
 Original & 89.77 & 90.42 & 82.98 \scriptsize(-7.44) & 80.66 & 81.64 & 75.43 \scriptsize(-6.21) &7.11 & 0.65& - & - \\
\midrule
Add Columns & 89.73 & 90.27 & 86.07 \scriptsize(-4.20) & 78.26 & 80.27 & 77.00 \scriptsize(-3.27) & 4.25 & 0.40& - & - \\
Remove Columns & 89.82 & 90.24 & 82.24 \scriptsize(-8.00)& 82.67 & 82.75 & 75.90 \scriptsize(-6.85) & 7.56 & 0.72& - & - \\ 
Remove Col in SQL & - & - & - & - & - & - & 5.02& -& - & 84.03\\
Rename Columns & 85.28 & 85.07 & 80.20 \scriptsize(-4.87) & 76.50 & 76.94 & 73.04 \scriptsize(-3.90) & 4.44 & 0.20 & - & -\\ 
\midrule
Add Tables & 57.88 & 89.50 & 88.78 \scriptsize(-0.72) & 63.81 & 81.14 & 80.71 \scriptsize(-0.37)& 0.33 & 0.07 & - & -\\ 
Remove Tables & - & - & - & - & - & - - & - & 1.62 & 83.86 & -\\
Rename Tables & 88.84 & 90.32 & 86.36 \scriptsize(-3.96) & 79.60 & 80.91 & 78.06 \scriptsize(-2.85) & 3.52 & 0.39 & - & -\\ 
Split Tables & 71.99 & 81.55 & 81.07 \scriptsize(-0.48) & 75.30 & 78.45 & 78.02 \scriptsize(-0.43) &  0.26 & 0.07 & - & -\\
Merge Tables & 85.29 & 87.03 & 82.18 \scriptsize(-5.15) & 65.56 & 67.09 & 63.59 \scriptsize(-3.50)  &  4.65 & 0.35 & - & -\\

\bottomrule
\end{tabular}
}
\end{table}

\subsection{Influence of Perturbation Types} 

\nop{\noindent\textbf{Table-level perturbation types are important for training the models to be robust on the major table-level schema evolution.}} We explore the effect of the column-level perturbation types and table-level perturbation types. As Table \ref{tab:perturbation_type_effect} shows, we train the model with both column-level and table-level perturbation types, and compare it with the model trained without column-level perturbation types and without table-level perturbation types. From our experiments, we found that without training on table-level perturbations, the model performance can be slightly better than the model trained with both column-level and table-level perturbation types on column-level perturbation types, while can lead to a significant performance drop on the table-level perturbation types. This indicates that the table-level perturbation data has a limited effect on the column-level perturbation types while having a huge impact on the table-level perturbation types. When looking at the model trained only on table-level perturbation types, we found that the model performance on both column-level and table-level perturbation types dropped. This indicates that the column-level perturbation types can still benefit the training.

\begin{table}[hb]
\centering
\caption{Irrelevant tables effect. ``w/”: the model is trained with all the perturbation types; ``w/o”: the model is only trained on the original training data; ``w/o+": the model is only trained on the original training data, but for the input table schema, we also add irrelevant tables.}
% \vspace{-0.5\baselineskip}
\label{tab:add_irrelevant_tables_effect}
\resizebox{0.7\linewidth}{!}{
\centering
\begin{tabular}{@{}l|ccc|ccc@{}}
\toprule
\multicolumn{7}{c}{Add Irrelevant Tables Effect} \\
\midrule
 \multirow{2}{*}{Perturbation Type} & \multicolumn{3}{c}{Table Match F1} & \multicolumn{3}{c}{Column Match F1} \\ 
 & w/o & w/o+ & w/ & w/o & w/o+ & w/\\
 \midrule
Original & 89.77 & 87.65 & \textbf{90.42} & 80.66 & 79.24 & \textbf{81.64} \\
\midrule
Add Columns & 89.73 & 86.35 & \textbf{90.27} & 78.26 & 75.31 & \textbf{80.27} \\
Remove Columns & 89.82 & 87.30 & \textbf{90.24} & 82.67 & 80.74 & \textbf{82.75}  \\ 
Rename Columns & \textbf{85.28} & 81.90 & 85.07 & 76.50 & 73.28 & \textbf{76.94}  \\ 
\midrule
Add Tables &  57.88 & 88.01 & \textbf{89.50} & 63.81 & 79.51 & \textbf{81.14}  \\ 
Remove Tables & - & - & - & - & - & -\\
Rename Tables & 88.84 & 86.84 & \textbf{90.32} & 79.60 & 78.47 & \textbf{80.91} \\ 
Split Tables &  71.99 & 67.27 & \textbf{81.55} & 75.30 & 70.39 & \textbf{78.45}  \\
Merge Tables & 85.29 & 83.56 & \textbf{87.03} & 65.56 & 63.59 & \textbf{67.09}\\

\bottomrule
\end{tabular}
}
\end{table}

% \vspace{-1em}

\vspace*{-0.5\baselineskip}
\subsection{Influence of Out-of-scope Types} 
\label{out_of_scope_exp}

We evaluate both in-scope and out-of-scope scenarios. In in-scope settings, schema changes may or may not alter the gold SQL. Out-of-scope cases involve two special perturbations: (1) \textit{Removing columns used in the gold SQL}, and (2) \textit{Removing tables used in the gold SQL}. In both cases, the schema lacks critical information, and the model is expected to abstain from generating a query.

To assess their impact, we train a model on a combined dataset that includes both out-of-scope and in-scope perturbation types, along with the original training data. We compare this model to others trained only on the original or in-scope data. As shown in Table~\ref{tab:out_of_scope_effect}, incorporating out-of-scope types results in performance degradation across both original and in-scope evaluation sets.

Error analysis reveals that the model trained with out-of-scope data tends to make more conservative predictions, sometimes abstaining even when the gold SQL is valid. Further analysis shows that the false positive (FP) rate closely matches the performance drop between models with and without out-of-scope training, confirming that increased conservatism is the main cause. Additionally, for the out-of-scope perturbations, the TP is only around 84\%, which indicates that the model still has a 16\% chance to make a prediction
even when there should not be an SQL.

\vspace*{-0.5\baselineskip}
\subsection{Influence of Irrelevant Tables}
We observed that the model trained with perturbation types demonstrates significant robustness to\nop{\input{ICLR 2025-text2sql/tables/add_irrelevant_tables}} table-level perturbations, such as adding and splitting tables. Upon analyzing the errors, we found that 
 models trained without perturbation types tend to predict SQL queries that join all available tables, even when some tables are irrelevant to the NLQs and SQLs. We hypothesize that this occurs because during training without perturbations, the model only sees relevant table schemas, causing it to learn spurious patterns that always try to join all the input tables.

To explore whether simply adding irrelevant tables could yield similar performance to models trained with perturbation data, we conducted an experiment where we trained CodeLlama on BIRD. As shown in Table \ref{tab:add_irrelevant_tables_effect}, adding irrelevant tables led to similar performance on "Add Tables" perturbation type. but it caused a performance drop on other perturbation types. This suggests that combining all perturbation data is necessary to train a more robust model.

% \footnotetext[6]{Since the schema evolution may lead to database restructure and there will be no existing values for the new database after schema change. We synthesize values to create new databases after schema evolution and execute the new gold SQLs after schema evolution on the new databases.}

\begin{table}[hb]
  \centering
\caption{Intra-database Effect. This experiment emphasizes that the training and evaluation occur within the same database, instead of across databases.}
% \vspace{-0.5\baselineskip}
\label{tab:intra_database_effect}
\small
\resizebox{0.6\linewidth}{!}{
\begin{tabular}{@{}l|cc|cc@{}}
\toprule
\multicolumn{5}{c}{Intra-database Effect} \\
\midrule
 \multirow{2}{*}{Perturbation Type} & \multicolumn{2}{c}{Table Match F1} & \multicolumn{2}{c}{Column Match F1} \\ 
 & w/o & w/ & w/o & w/\\
 \midrule
Original &  87.24 & \textbf{87.43} & 79.54 & \textbf{80.89}\\
\midrule
Add Columns & 87.14 & \textbf{87.43} & 76.36 & \textbf{78.92}\\
Remove Columns & \textbf{87.29} & 87.27 & 81.14 & \textbf{81.29} \\ 
Rename Columns & 85.71 & \textbf{86.43} & 77.45 & \textbf{79.09} \\ 
\midrule
Add Tables & 61.13 & \textbf{83.95} & 66.11 & \textbf{78.57}  \\ 
Remove Tables & - & - & - & -\\
Rename Tables & 86.33 & \textbf{86.67} & 79.44 & \textbf{79.96}\\ 
Split Tables & 71.82 & \textbf{78.52} & 75.09 & \textbf{77.42}  \\
Merge Tables & 85.11 & \textbf{87.44} & 71.43 & \textbf{74.72}\\

\bottomrule
\end{tabular}
}
\end{table}
\vspace*{-2.0\baselineskip}
\subsection{Influence of Intra-DB and Cross-DB}
We hypothesize that a model trained on the same databases may not only learn schema evolution patterns but also become familiar with specific table and column names. To test this, we split the BIRD training data into train/test sets to ensure that each database in the test set also appears in the training set. We use Code Llama as the base model. The results in Table \ref{tab:intra_database_effect} show that, for most perturbation types, the model's performance improves more compared to the cross-database scenario in Section \ref{main_results}, which verifies our hypothesis.

\nop{\subsection{Generalizability to Other Datasets}
To evaluate the generalizability of \texttt{EvoSchema} to other text-to-SQL datasets, we conducted experiments on the Spider dataset and used Mistral as the base model. As shown in Table \ref{tab:spider_result}, we reached conclusions consistent with those in Section \ref{main_results}, which further demonstrates the effectiveness and utility of our proposed framework and training methods.}

% \input{ICLR 2025-text2sql/tables/intra_database_effect}

% \input{ICLR 2025-text2sql/tables/spider_results}

% \subsection{results on other datasets}
% \input{ICLR 2025-text2sql/tables/spider_main_results}

\section{Conclusion}

In conclusion, we formulate the critical challenge of schema evolution in adaptive text-to-SQL systems and introduce \texttt{EvoSchema}, a comprehensive, diverse and unique benchmark designed specifically to study and address this problem. We developed a structured taxonomy of schema evolution types, enabling the synthesis of realistic schema designs through column-level and table-level perturbations. Using this taxonomy, we construct an evaluation benchmark to rigorously assess model robustness under schema changes and also introduce a novel training paradigm that augments existing \textit{\textless NLQ, relevant schema, SQL\textgreater} triples with diverse schema designs for training to improve robustness against schema evolution.

\nop{Moreover, \texttt{EvoSchema} goes beyond evaluation by inspiring a novel training paradigm. By augmenting existing \textit{\textless NLQ, relevant schema, SQL\textgreater} triples with a variety of schema designs, our benchmark introduces greater data diversity and compels models to recognize schema differences during training. As a result, text-to-SQL models trained with \texttt{EvoSchema} achieve significant improvements, with up to a 33-point gain in performance on schema perturbation benchmarks compared to models trained on unperturbed data. These results underscore the pivotal role of \texttt{EvoSchema} in advancing the robustness and adaptability of text-to-SQL systems, providing a solid foundation for future research and innovation in addressing schema evolution challenges.}

\begin{acks}
 The authors would like to thank colleagues from
the OSU NLP group for their insightful discussions and constructive suggestions
and all anonymous reviewers for their thoughtful comments. 
\end{acks}

\nop{
\section{Introduction}

Lorem ipsum dolor sit amet, consectetur adipiscing elit. Suspendisse a arcu quis arcu malesuada ultricies vitae in felis. Curabitur porta lacus at felis viverra hendrerit in non eros. Nam tempus tincidunt metus vitae fermentum. Donec sed risus felis. Cras luctus massa elementum, semper urna vel, efficitur ipsum. Morbi at tellus libero.

Praesent imperdiet, lacus nec varius placerat, est ex eleifend justo, a vulputate leo massa consectetur nunc. Donec posuere in mi ut tempus. Pellentesque sem odio, faucibus non mi in, laoreet maximus arcu. In hac habitasse platea dictumst. Nunc euismod neque eu urna accumsan, vitae vehicula metus tincidunt. Maecenas congue tortor nec varius pellentesque. Pellentesque bibendum libero ac dignissim euismod. Aliquam justo ante, pretium vel mollis sed, consectetur accumsan nibh. Nulla sit amet sollicitudin est. 

\section{Core Structural Elements}

Nulla placerat feugiat augue, id blandit urna pretium nec. Nulla velit sem, tempor vel mauris ut, porta commodo quam. Donec lectus erat, sodales eu mauris eu, fringilla vestibulum nisl. Morbi viverra tellus id lorem faucibus cursus. Quisque et orci in est faucibus semper vel a turpis. Vivamus posuere sed ligula et. 

\subsection{Figures}

Aliquam justo ante, pretium vel mollis sed, consectetur accumsan nibh. Nulla sit amet sollicitudin est. Etiam ullamcorper diam a sapien lacinia faucibus. Duis vulputate, nisl nec tincidunt volutpat, erat orci eleifend diam, eget semper risus est eget nisl. Donec non odio id neque pharetra ultrices sit amet id purus. Nulla non dictum tellus, id ullamcorper libero. Curabitur vitae nulla dapibus, ornare dolor in, efficitur enim. Cras fermentum facilisis elit vitae egestas. Nam vulputate est non tellus efficitur pharetra. Vestibulum ligula est, varius in suscipit vel, porttitor id massa. Nulla placerat feugiat augue, id blandit urna pretium nec. Nulla velit sem, tempor vel mauris ut, porta commodo quam \autoref{fig:duck}.

\begin{figure}
  \centering
  \includegraphics[width=\linewidth]{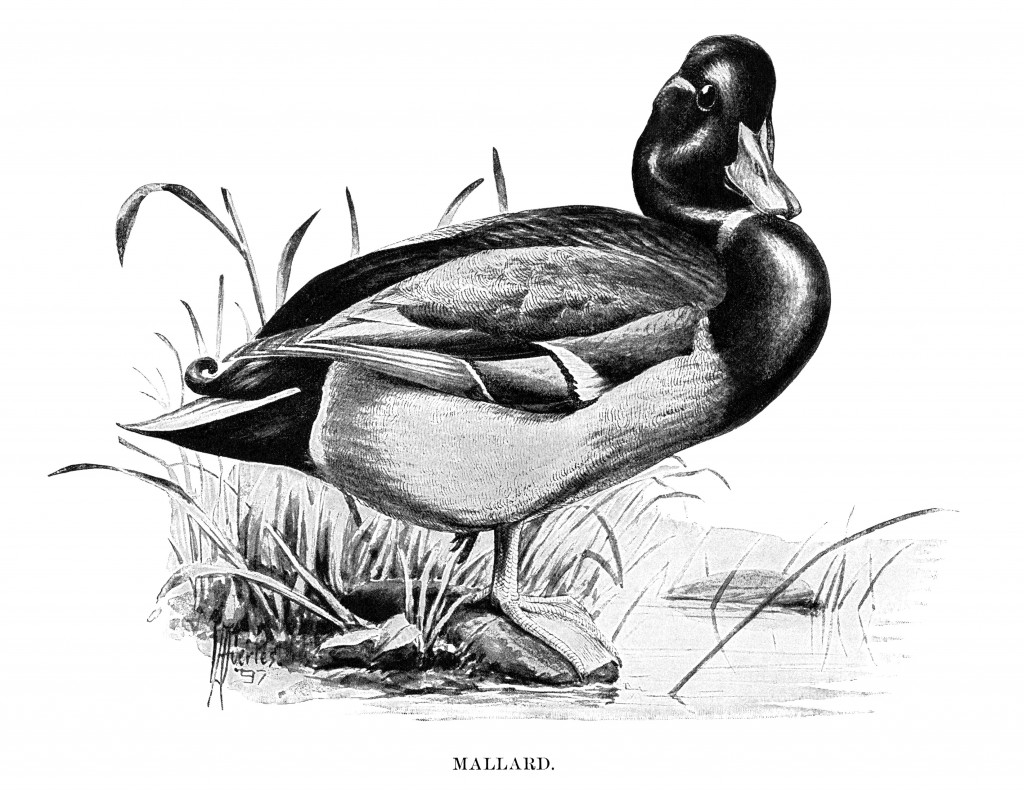}
  \caption{An illustration of a Mallard Duck. Picture from Mabel Osgood Wright, \textit{Birdcraft}, published 1897.}
  \label{fig:duck}
\end{figure}

\begin{table*}[t]
  \caption{A double column table.}
  \label{tab:commands}
  \begin{tabular}{ccl}
    \toprule
    A Wide Command Column & A Random Number & Comments\\
    \midrule
    \verb|\tabular| & 100& The content of a table \\
    \verb|\table|  & 300 & For floating tables within a single column\\
    \verb|\table*| & 400 & For wider floating tables that span two columns\\
    \bottomrule
  \end{tabular}
\end{table*}

\subsection{Tables}

Curabitur vitae nulla dapibus, ornare dolor in, efficitur enim. Cras fermentum facilisis elit vitae egestas. Mauris porta, neque non rutrum efficitur, odio odio faucibus tortor, vitae imperdiet metus quam vitae eros. Proin porta dictum accumsan \autoref{tab:commands}.

Duis cursus maximus facilisis. Integer euismod, purus et condimentum suscipit, augue turpis euismod libero, ac porttitor tellus neque eu enim. Nam vulputate est non tellus efficitur pharetra. Aenean molestie tristique venenatis. Nam congue pulvinar vehicula. Duis lacinia mollis purus, ac aliquet arcu dignissim ac \autoref{tab:freq}. 

\begin{table}[hb]% h asks to places the floating element [h]ere.
  \caption{Frequency of Special Characters}
  \label{tab:freq}
  \begin{tabular}{ccl}
    \toprule
    Non-English or Math & Frequency & Comments\\
    \midrule
    \O & 1 in 1000& For Swedish names\\
    $\pi$ & 1 in 5 & Common in math\\
    \$ & 4 in 5 & Used in business\\
    $\Psi^2_1$ & 1 in 40\,000 & Unexplained usage\\
  \bottomrule
\end{tabular}
\end{table}

Nulla sit amet enim tortor. Ut non felis lectus. Aenean quis felis faucibus, efficitur magna vitae. Curabitur ut mauris vel augue tempor suscipit eget eget lacus. Sed pulvinar lobortis dictum. Aliquam dapibus a velit.

\subsection{Listings and Styles}

Aenean malesuada fringilla felis, vel hendrerit enim feugiat et. Proin dictum ante nec tortor bibendum viverra. Curabitur non nibh ut mauris egestas ultrices consequat non odio.

\begin{itemize}
\item Duis lacinia mollis purus, ac aliquet arcu dignissim ac. Vivamus accumsan sollicitudin dui, sed porta sem consequat.
\item Curabitur ut mauris vel augue tempor suscipit eget eget lacus. Sed pulvinar lobortis dictum. Aliquam dapibus a velit.
\item Curabitur vitae nulla dapibus, ornare dolor in, efficitur enim.
\end{itemize}

Ut sagittis, massa nec rhoncus dignissim, urna ipsum vestibulum odio, ac dapibus massa lorem a dui. Nulla sit amet enim tortor. Ut non felis lectus. Aenean quis felis faucibus, efficitur magna vitae. 

\begin{enumerate}
\item Duis lacinia mollis purus, ac aliquet arcu dignissim ac. Vivamus accumsan sollicitudin dui, sed porta sem consequat.
\item Curabitur ut mauris vel augue tempor suscipit eget eget lacus. Sed pulvinar lobortis dictum. Aliquam dapibus a velit.
\item Curabitur vitae nulla dapibus, ornare dolor in, efficitur enim.
\end{enumerate}

Cras fermentum facilisis elit vitae egestas. Mauris porta, neque non rutrum efficitur, odio odio faucibus tortor, vitae imperdiet metus quam vitae eros. Proin porta dictum accumsan. Aliquam dapibus a velit. Curabitur vitae nulla dapibus, ornare dolor in, efficitur enim. Ut maximus mi id arcu ultricies feugiat. Phasellus facilisis purus ac ipsum varius bibendum.

\subsection{Math and Equations}

Curabitur vitae nulla dapibus, ornare dolor in, efficitur enim. Cras fermentum facilisis elit vitae egestas. Nam vulputate est non tellus efficitur pharetra. Vestibulum ligula est, varius in suscipit vel, porttitor id massa. Cras facilisis suscipit orci, ac tincidunt erat.
\begin{equation}
  \lim_{n\rightarrow \infty}x=0
\end{equation}

Sed pulvinar lobortis dictum. Aliquam dapibus a velit porttitor ultrices. Ut maximus mi id arcu ultricies feugiat. Phasellus facilisis purus ac ipsum varius bibendum. Aenean a quam at massa efficitur tincidunt facilisis sit amet felis. 
\begin{displaymath}
  \sum_{i=0}^{\infty} x + 1
\end{displaymath}

Suspendisse molestie ultricies tincidunt. Praesent metus ex, tempus quis gravida nec, consequat id arcu. Donec maximus fermentum nulla quis maximus.
\begin{equation}
  \sum_{i=0}^{\infty}x_i=\int_{0}^{\pi+2} f
\end{equation}

Curabitur vitae nulla dapibus, ornare dolor in, efficitur enim. Cras fermentum facilisis elit vitae egestas. Nam vulputate est non tellus efficitur pharetra. Vestibulum ligula est, varius in suscipit vel, porttitor id massa. Cras facilisis suscipit orci, ac tincidunt erat.

\section{Citations}

Some examples of references. A paginated journal article~\cite{Abril07}, an enumerated journal article~\cite{Cohen07}, a reference to an entire issue~\cite{JCohen96}, a monograph (whole book) ~\cite{Kosiur01}, a monograph/whole book in a series (see 2a in spec. document)~\cite{Harel79}, a divisible-book such as an anthology or compilation~\cite{Editor00} followed by the same example, however we only output the series if the volume number is given~\cite{Editor00a} (so Editor00a's series should NOT be present since it has no vol. no.), a chapter in a divisible book~\cite{Spector90}, a chapter in a divisible book in a series~\cite{Douglass98}, a multi-volume work as book~\cite{Knuth97}, an article in a proceedings (of a conference, symposium, workshop for example) (paginated proceedings article)~\cite{Andler79}, a proceedings article with all possible elements~\cite{Smith10}, an example of an enumerated proceedings article~\cite{VanGundy07}, an informally published work~\cite{Harel78}, a doctoral dissertation~\cite{Clarkson85}, a master's thesis~\cite{anisi03}, an finally two online documents or world wide web resources~\cite{Thornburg01, Ablamowicz07}.

\begin{acks}
 This work was supported by the [...] Research Fund of [...] (Number [...]). Additional funding was provided by [...] and [...]. We also thank [...] for contributing [...].
\end{acks}

}

%\clearpage

\bibliographystyle{ACM-Reference-Format}
\bibliography{sample}

\end{document}